\documentclass[aps,prd,epsf,showpacs]{revtex4-2}
	\usepackage[pdftex]{graphicx}

\usepackage{amsmath}
\usepackage{amsfonts}
\usepackage{amssymb}
\usepackage{dsfont}
\usepackage[]{units}
\usepackage{braket}
\usepackage[utf8]{inputenc}
\usepackage[T1]{fontenc}
\usepackage{url}
\usepackage[english]{babel}
\usepackage{bm}
\usepackage{url}
\usepackage{verbatim}
\usepackage[]{cleveref}
\usepackage{bm}
\usepackage{slashed}
\usepackage{array}
\usepackage{algorithm}
\usepackage{float}
\usepackage{esvect}
\usepackage{bbold}
\usepackage{comment}
\usepackage{color, colortbl}
\usepackage{xcolor}
\usepackage{setspace}
\usepackage{cancel}

\newcommand{\be}{\begin{equation}}
\newcommand{\ee}{\end{equation}}
\newcommand{\beq}{\begin{equation}}
\newcommand{\eeq}{\end{equation}}
\newcommand{\bea}{\begin{eqnarray}}
\newcommand{\eea}{\end{eqnarray}}

\newcommand{\bie}{\begin{itemize}}
\newcommand{\ie}{\item}
\newcommand{\eie}{\end{itemize}}
\newcommand{\ben}{\begin{enumerate}}
\newcommand{\een}{\end{enumerate}}
\newcommand{\f}[1]{\textbf{#1}}
\newcommand{\J}{{\tilde{J}}}

\newcommand{\VQQ}{V_{\bar{Q}Q}(r)}

\newcommand{\MatrixThreeXThree}[9]{	\left( \begin{array}{ccc} #1 & #2 & #3 \\ 
		#4 & #5 & #6 \\ 
		#7 & #8 & #9 \\\end{array}\right)}

\newcommand{\gtapprox}{\raisebox{-0.5ex}{$\,\stackrel{>}{\scriptstyle\sim}\,$}}
\newcommand{\ltapprox}{\raisebox{-0.5ex}{$\,\stackrel{<}{\scriptstyle\sim}\,$}}

\definecolor{darkred}{rgb}{0.4,0.0,0.0}
\definecolor{darkgreen}{rgb}{0.0,0.4,0.0}
\definecolor{darkblue}{rgb}{0.0,0.0,0.4}

\begin{document}

\title{ 
Study of $I=0$ bottomonium bound states and resonances in $S$, $P$, $D$ and $F$ waves with lattice QCD static-static-light-light potentials 
}

\author{$^{(1)}$Pedro Bicudo}
\email{bicudo@tecnico.ulisboa.pt}

\author{$^{(1)}$Nuno Cardoso}
\email{nuno.cardoso@tecnico.ulisboa.pt}

\author{$^{(2)}$Lasse Mueller}
\email{lmueller@itp.uni-frankfurt.de}

\author{$^{(2),(3)}$Marc Wagner}
\email{mwagner@itp.uni-frankfurt.de}

\affiliation{\vspace{0.1cm}$^{(1)}$CeFEMA, Dep.\ F\'{\i}sica, Instituto Superior T\'ecnico, Universidade de Lisboa, Av.\ Rovisco Pais, 1049-001 Lisboa, Portugal}

\affiliation{\vspace{0.1cm}$^{(2)}$Johann Wolfgang Goethe-Universit\"at Frankfurt am Main, Institut f\"ur Theoretische Physik, Max-von-Laue-Stra{\ss}e 1, D-60438 Frankfurt am Main, Germany}

\affiliation{\vspace{0.1cm}$^{(3)}$Helmholtz Research Academy Hesse for FAIR, Campus Riedberg, \\ Max-von-Laue-Stra{\ss}e 12, D-60438 Frankfurt am Main, Germany}

\begin{abstract}

In this paper we study $I = 0$ bottomonium in $S$, $P$, $D$ and $F$ waves considering five coupled channels, one confined quarkonium and four open $B^{(*)} \bar B^{(*)}$ and $B_s^{(*)} \bar B_s^{(*)}$ meson-meson channels.
To this end we use and extend a recently developed novel approach utilizing lattice QCD string breaking potentials for the study of quarkonium bound states and resonances. This approach is based on the Born Oppenheimer approximation and the unitary emergent wave method and allows to compute the poles of the $\mbox{T}$ matrix.
We compare our results to existing experimental results for $I = 0$ bottomonium and discuss masses, decay widths and the assignment of angular momentum quantum numbers. Moreover, we determine the quarkonium and meson-meson composition of these states to clarify, which of them are ordinary quarkonium, and which of them should rather be interpreted as tetraquarks.

\end{abstract}

\maketitle

\section{Introduction\label{sec:intro}}

In the last decade a whole new class of hadrons has been discovered experimentally, so-called tetraquarks, which are composed of two quarks and two antiquarks \cite{LHCb:2021uow,BESIII:2020qkh,LHCb:2020bls,LHCb:2020pxc,JPAC:2018zyd,LHCb:2018oeg,LHCb:2015sqg,BESIII:2015pqw,Belle:2014nuw,LHCb:2014zfx,BESIII:2013qmu,BESIII:2013ouc,BESIII:2013mhi,Belle:2013shl,Xiao:2013iha,Belle:2013yex,BESIII:2013ris,Ketzer:2012vn,Belle:2011aa,Belle:2009lvn,Belle:2007hrb}. In particular, a larger number of such tetraquarks observed in the last couple of years at Belle, BESIII and LHCb have at least one heavy quark, as anticipated in Refs.\ \cite{Ader:1981db,Ballot:1983iv}. From the onset of QCD the existence of tetraquarks was expected \cite{Jaffe:1976ig}, However, a quantitative first principles prediction of their properties, e.g.\ quark composition, quantum numbers, masses and decay widths, using e.g.\ lattice QCD, remains to be achieved. The main reason, why this has not been successful yet, is that the majority of observed heavy tetraquarks are resonances high in the quarkonium spectrum (the only exception is the $T_{cc}$ tetraquark recently found by LHCb at CERN \cite{LHCb:2021vvq,LHCb:2021auc}). Studying such resonances with lattice QCD is possible in principle using the L\"uscher phase shift method \cite{Luscher:1990ux}, but practically feasible only for a single or a small number of decay channels. For several open channels, as it is the case for some of the recently observed heavy tetraquarks, following this path seems tremendously difficult.

Because of these difficulties we recently started to develop another approach \cite{Bicudo:2019ymo,Bicudo:2020qhp}, utilizing lattice QCD results for static potentials, to study bottomonium as well as tetraquark resonances with the same non-exotic quantum numbers high in the spectrum. We start with lattice QCD potentials computed with static quarks and light quarks in the context of string breaking \cite{Bali:2005fu,Bulava:2019iut,Bonati:2020orj}, which provide information on the interactions between a two-quark quarkonium channel and several four-quark meson-meson channels. We use the Born-Oppenheimer diabatic approximation as in Refs.\ \cite{Bicudo:2012qt,Brown:2012tm}, i.e.\ include the kinetic energy of the heavy quarks, and study the dynamically coupled quarkonium and meson-meson channels with techniques from quantum mechanics.

In our previous work \cite{Bicudo:2019ymo,Bicudo:2020qhp} we applied this method to systems with an $S$ wave bottonomium channel coupled to a $B^{(*)} \bar B^{(*)}$ and a $B_s^{(*)} \bar B_s^{(*)}$ channel. In this work we extend our studies to $P$ wave, $D$ wave and $F$ wave quarkonium channels, again coupled to $B^{(*)} \bar B^{(*)}$ and $B_s^{(*)} \bar B_s^{(*)}$ channels. Because of the intrinsic parity of quarks, the relative orbital angular momentum of our meson-meson channels differs by one unit from the quarkonium orbital angular momentum. Compared to the $S$ wave case, this doubles the number of meson-meson channels, and, thus, we have to take into account five coupled channels. As mentioned in the previous paragraph, such a large number of coupled channels seems currently inaccessible for fully dynamical lattice QCD studies of resonances.

Using our approach we were already able to address claims in the literature, from studies with different hadronic models, on the nature of some of the bottonomium resonances observed at Belle. For example, it has been discussed, which of the observed excited bottonomium resonances are $S$ wave or $D$ wave states \cite{Li:2019qsg,Liang:2019geg,Giron:2020qpb}, since the total angular momentum is not yet experimentally determined, and whether they have a large meson-meson content \cite{Meng:2007tk,Simonov:2008ci,Voloshin:2012dk,Sungu:2018iew}. Of particular interest is the nature of the newly discovered resonance $\Upsilon(10753)$ recently observed at Belle with mass $(10.753 \pm 0.007) \, \text{GeV}$ \cite{Belle:2019cbt}. Model and effective field theory calculations suggest for instance this resonance to be either a tetraquark \cite{Wang:2019veq,Ali:2019okl}, a hybrid meson \cite{TarrusCastella:2019lyq,Chen:2019uzm,Brambilla:2019esw} or the more canonical and so far missing $\Upsilon(3D)$ \cite{Li:2019qsg,Liang:2019geg,Giron:2020qpb}. With our lattice QCD based approach we found a pole of the $\mbox{T}$ matrix in the $S$ wave channel corresponding to the mass $10.773 \, \text{GeV}$ \cite{Bicudo:2020qhp}, very similar to the Belle measurement. Moreover, we found a large meson-meson component for that state. Thus, we proposed that the recently observed $\Upsilon(10753)$ is a dynamical state, composed mostly of a meson-meson pair.

By extending our work to $P$ wave, $D$ wave and $F$ wave channels, we can now check, whether there is also a candidate for the $\Upsilon(10753)$ in the $D$ wave channel. Moreover, we will obtain a complete picture of the $I = 0$ bottomonium spectrum. In particular we can identify states with a large meson-meson component, which can be interpreted as tetraquarks. Finally, it is interesting to explore the possible existence of a bottomonium counterpart of $X(3872)$ discovered at Belle and CDF \cite{Belle:2003nnu,CDF:2003cab}, which is extremely close to the $D \bar D^*$ and $D^* \bar D$ thresholds. In our case, this would correspond to a resonance very close to either the $B^{(*)} \bar B^{(*)}$ or the $B_s^{(*)} \bar B_s^{(*)}$ thresholds.

This paper is organized as follows. In Section~\ref{SEC002} we detail our approach, where we use lattice QCD potentials computed for string breaking in a coupled channel Schrödinger equation. In particular we carry out a partial wave decomposition and derive $5 \times 5$ Schrödinger equations for the $P$ wave, $D$ wave and $F$ wave channels. Using the emergent wave method, we also show, how the $\mbox{T}$ matrix can be computed, leading to both the phase shifts and, after a pole search, to masses and decay widths. In Section~\ref{SEC003} we show and discuss corresponding numerical results. We also determine the quarkonium and meson-meson  composition for all states. Finally, in Section~\ref{SEC004}, we conclude on the points raised in this introduction and discuss possibilities for future research within our formalism.

\begin{widetext}

\section{\label{SEC002}Quarkonium resonances from lattice QCD static potentials}

\subsection{Quantum numbers}

We consider a heavy quark-antiquark pair and either no light quarks ($\bar Q Q$) or a light quark-antiquark pair with isospin $I = 0$ ($\bar Q Q (\bar u u + \bar d d) \equiv \bar M M$ or $\bar Q Q \bar s s \equiv \bar M_s M_s$). In the limit, where the heavy quarks are static, such a system can be characterized by the following quantum numbers:
\begin{itemize}
	\item $J^{PC}$: total angular momentum, parity and charge conjugation.
	\item $S^{PC}_Q$: spin of the heavy quark-antiquark pair and corresponding parity and charge conjugation.
	\item $\J^{PC}$: total angular momentum excluding the heavy quark spins and corresponding parity and charge conjugation. (For quarkonium $\J^{PC}$ coincides with the orbital angular momentum $L^{PC}$ of the heavy quark-antiquark pair.)
\end{itemize}
Moreover, the majority of observables, in particular energy levels, do not depend on $S^{PC}_Q$. Thus, the relevant quantum numbers in our context are $\J^{PC}$ and not, as usual, $J^{PC}$.

For rather heavy $b$ quarks we expect that our approach, which is based on static symmetries and quantum numbers, will yield reasonably accurate results, possibly even for $c$ quarks.

\subsection{\label{SEC488}The coupled channel Schrödinger equation}

Now we discuss the coupled channel Schrödinger equation, which we have derived in detail in our previous papers \cite{Bicudo:2019ymo,Bicudo:2020qhp}. We take the two lowest meson decay channels into account, where each channel contains two negative parity heavy-light mesons, either $\bar M M$ or $\bar M_s M_s$. The corresponding light spin is $S_q^{PC} = 1^{--}$ \cite{Bicudo:2019ymo}. In addition to these decay channels there is, of course, also the quarkonium channel $\bar Q Q$. This amounts to a seven-component wave function $\psi(\f{r}) = (\psi_{\bar{Q} Q}(\f{r}), \vec{\psi}_{\bar{M} M}(\f{r}), \vec{\psi}_{\bar{M}_s M_s}(\f{r}))$, where the first component represents the $\bar{Q} Q$ channel and the remaining six components the spin-1 triplets of the $\bar M M$ and the $\bar M_s M_s$ channel, respectively. $\f{r}$ denotes the relative coordinates of the heavy quark-antiquark pair.

The coupled channel Schrödinger equation reads
\begin{align}
\Bigg(-\frac{1}{2} \mu^{-1} \bigg(\partial_r^2 + \frac{2}{r} \partial_r - \frac{\f{L}^2}{r^2}\bigg) + V(\f{r}) + \MatrixThreeXThree{E_{\text{threshold}}}{0}{0}{0}{2m_M}{0}{0}{0}{2m_{M_s}}-E\Bigg) \psi(\f{r}) = 0 , \label{eqn:schroedinger_equation}
\end{align}
where $\mu^{-1} = \textrm{diag}(1/\mu_Q, 1/\mu_M, 1/\mu_M, 1/\mu_M, 1/\mu_{M_s}, 1/\mu_{M_s}, 1/\mu_{M_s})$ is a $7 \times 7$ diagonal matrix with the reduced masses of the heavy quarks and the heavy mesons, $\mu_Q = m_Q/2$, $\mu_M = m_M/2$ and $\mu_{M_s} = m_{M_s}/2$. In the static limit the pseudoscalar and the vector heavy-light meson masses are identical. For finite heavy quark mass there is a small difference in these meson masses, e.g.\ for heavy $b$ quarks $m_{B^*} - m_B \approx 45 \, \text{MeV}$ and $m_{B_s^*} - m_{B_s} = 49 \, \text{MeV}$. We take the spin averaged masses for $m_M$ and $m_{M_s}$ (see Section~\ref{SEC433}). $\f{L} = \f{r} \times \f{p}$ denotes the orbital angular momentum operator and $E_{\text{threshold}}$ is the threshold energy corresponding to two negative parity static-light mesons in the same lattice setup, where the static potentials are computed (for more details see Section~\ref{SEC433} and Ref.\ \cite{Bicudo:2020qhp}). The potential matrix $V(\f{r})$ is given by
\begin{align}
	V(\f{r}) = \MatrixThreeXThree{\VQQ}{V_{\textrm{mix}}(r)\left(1\otimes\f{e}_r\right)}{(1/\sqrt{2}) V_{\textrm{mix}}(r)\left(1\otimes\f{e}_r\right)}
	{V_{\textrm{mix}}(r)\left(\f{e}_r\otimes 1\right)}{V_{\bar{M}M}(r)}{0}
	{(1/\sqrt{2}) V_{\textrm{mix}}(r)\left(\f{e}_r\otimes 1\right)}{0}{V_{\bar{M}M}(r)}
\end{align}
with
\begin{align}
V_{\bar{M}M}(r) = V_{\bar{M}M, \parallel}(r) \Big(\f{e}_r\otimes\f{e}_r\Big) + V_{\bar{M}M, \perp}(r) \Big(1-\f{e}_r\otimes\f{e}_r\Big) ,
\end{align}
where we have assumed that meson-meson interactions vanish. $\VQQ$, $V_{\bar{M}M, \parallel}(r)$, $V_{\bar{M}M, \perp}(r)$ and $V_{\textrm{mix}}(r)$ can be expressed in terms of QCD static potentials, which can be computed with lattice QCD (see Refs.\ \cite{Bali:2005fu,Bicudo:2019ymo} for details). $\VQQ$ represents the potential of a heavy quark-antiquark pair, $V_{\bar{M}M, \parallel}(r)$, $V_{\bar{M}M, \perp}(r)$ the interaction between a pair of heavy-light mesons and $V_{\textrm{mix}}(r)$ describes the mixing of the quarkonium channel and the meson-meson channels. We use the same potentials both for the $\bar M M$ channel and the $\bar M_s M_s$ channel and expect this to be reasonable, because the light quark dependence of static potentials is known to be rather mild. We confirmed this expectation by carrying out a consistency check with a recent $2+1$-flavor lattice study of string breaking \cite{Bulava:2019iut} (for more details see our previous work \cite{Bicudo:2020qhp}). Note that, the mixing between the $\bar Q Q$ channel and the $\bar M_s M_s$ channel is suppressed by $1/\sqrt{2}$, because there is only a single strange quark flavor in comparison to the two degenerate light flavors corresponding to the $\bar M M$ channel.  

We use lattice QCD data from Ref.\ \cite{Bali:2005fu} to determine continuous functions $\VQQ$, $V_{\bar{M} M, \parallel}(r)$, $V_{\bar{M} M, \perp}(r)$ and $V_{\textrm{mix}}(r)$. The data points for $\VQQ$ and $V_{\textrm{mix}}(r)$ are consistently parameterized by
\begin{align}
	& \VQQ = E_0 - \frac{\alpha}{r} + \sigma r + \sum_{j=1}^{2} c_{\bar{Q}Q, j} \, r \exp\bigg(-\frac{r^2}{2\lambda^2_{\bar{Q}Q, j}}\bigg) \label{eqn:parameterization1} \\
	& V_{\textrm{mix}}(r) = \sum_{j=1}^{2} c_{\textrm{mix}, j} \, r \exp\bigg(-\frac{r^2}{2\lambda^2_{\textrm{mix}, j}}\bigg)
	\label{eqn:parameterization3}
\end{align}
with parameters $E_0$, $\alpha$, $\sigma$, $c_{\bar{Q}Q, j}$, $\lambda_{\bar{Q}Q, j}$, $c_{\textrm{mix}, j}$, $\lambda_{\textrm{mix}, j}$ collected in Table~\ref{tab:parameters}. The data points for $V_{\bar{M} M, \parallel}(r)$ are consistent with
\begin{align}
V_{\bar{M} M, \parallel}(r) = 0 . \label{eqn:parameterization2}
\end{align}
For $V_{\bar{M} M, \perp}(r)$ no lattice data is available yet. We assume 
\begin{align}
V_{\bar{M} M, \perp}(r) = 0
\end{align}
(see also Ref.\ \cite{Bicudo:2019ymo}). In Fig.\ \ref{fig:parametrizations} we show the data points for $\VQQ$, $V_{\bar{M}M, \parallel}(r)$ and $V_{\textrm{mix}}(r)$ together with the parameterizations (\ref{eqn:parameterization1}) to (\ref{eqn:parameterization2}).

\begin{table}[htb]
	\centering
	\begin{tabular}{c|c|c}
		\hline
		potential & parameter & value \\ 
		\hline
		\hline
		$V_{\bar{Q} Q}(r)$ & $E_0$                      & $-1.599(269) \, \textrm{GeV}\phantom{1.^{-1}}$ \\
		& $\alpha$                   & $+0.320(94) \phantom{1.0 \, \textrm{GeV}^{-1}}$ \\
		& $\sigma$                   & $+0.253(035) \, \textrm{GeV}^{2\phantom{-}}\phantom{1.}$ \\
		& $c_{\bar{Q} Q,1}$          & $+0.826(882) \, \textrm{GeV}^{2\phantom{-}}\phantom{1.}$ \\
		& $\lambda_{\bar{Q} Q,1}$    & $+0.964(47) \, \textrm{GeV}^{-1}\phantom{1.0}$ \\
		& $c_{\bar{Q} Q,2}$          & $+0.174(1.004) \, \textrm{GeV}^{2\phantom{-}}$ \\
		& $\lambda_{\bar{Q} Q,2}$    & $+2.663(425) \, \textrm{GeV}^{-1}\phantom{1.}$ \\
		\hline
		\hline
		$V_{\textrm{mix}}(r)$ & $c_{\textrm{mix},1}$       & $-0.988(32) \, \textrm{GeV}^{2\phantom{-}}\phantom{1.0}$ \\
		& $\lambda_{\textrm{mix},1}$ & $+0.982(18) \, \textrm{GeV}^{-1}\phantom{1.0}$ \\
		& $c_{\textrm{mix},2}$       & $-0.142(7) \, \textrm{GeV}^{2\phantom{-}}\phantom{1.00}$ \\

		& $\lambda_{\textrm{mix},2}$ & $+2.666(46) \, \textrm{GeV}^{-1}\phantom{1.0}$ \\
		\hline
	\end{tabular}
	\caption{Parameters of the potential parameterizations \eqref{eqn:parameterization1} and \eqref{eqn:parameterization3}.}
	\label{tab:parameters}
\end{table}

\begin{figure}[htb]
	\includegraphics[width=0.4\textwidth]{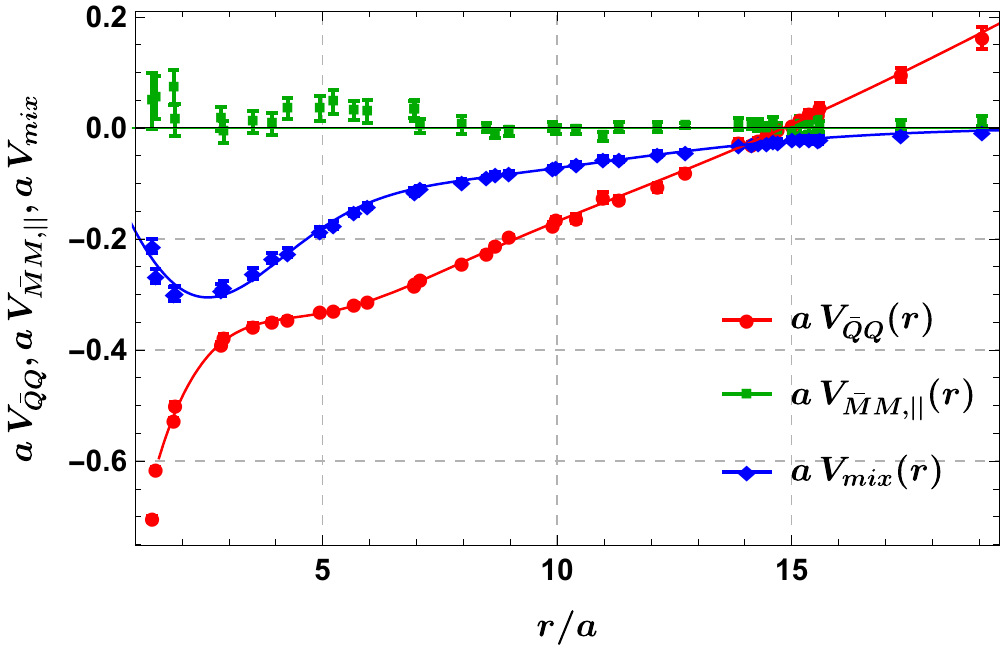}
	\caption{\label{fig:parametrizations}Data points for $\VQQ$, $V_{\bar{M}M, \parallel}(r)$ and $V_{\textrm{mix}}(r)$. The curves represent the parameterizations \eqref{eqn:parameterization1} to \eqref{eqn:parameterization2}.}
\end{figure}

\subsection{\label{SEC390}The coupled channel Schrödinger equation for total angular momentum $\J$}

Now we specialize the coupled channel Schrödinger equation (\ref{eqn:schroedinger_equation}) to study quarkonium bound states and resonances and meson-meson scattering for arbitrary given $\J$. For details we refer to Section~II~D of Ref.\ \cite{Bicudo:2019ymo}, where the mathematical formalism is discussed for $\J = 0$.

At first, each of the meson-meson components of the wave function $\psi(\f{r})$ is written as sum of an incident wave, which is a solution of the free Schrödinger equation, i.e.\ Eq.\ (\ref{eqn:schroedinger_equation}) with $V(\f{r}) = 0$, and an emergent wave. This will allow us to define and determine scattering amplitudes for energies above the lowest meson-meson threshold. In contrast to our previous work \cite{Bicudo:2019ymo}, we do not restrict the scattering problem to an incident plane wave, but allow arbitrary regular solutions of the free Schrödinger equation, including superpositions of $\bar M M$ and $\bar M_s M_s$ waves. We expand $\psi(\f{r}) = (\psi_{\bar{Q} Q}(\f{r}), \vec{\psi}_{\bar{M} M}(\f{r}), \vec{\psi}_{\bar{M}_s M_s}(\f{r}))$ in terms of eigenfunctions of $\tilde{\f{J}}^2$ and $\J_z$,
\begin{align}
	\psi_{QQ}(\f{r}) &= \frac{u_{0,0}(r)}{r}Y_{0,0}(\Omega) + \sum_{\bar{J}=1}^{\infty} \sum_{\J_z=-\J}^{+\J} \frac{u_{\J, \J_z}(r)}{r}Y_{\J, \J_z}(\Omega)\\
	\vec{\psi}_{\bar{M}_{(s)} M_{(s)}}(\f{r}) &= \alpha_{\bar{M}_{(s)} M_{(s)},0,0}\, j_0(k_{(s)}r)\,\f{Z}_{1\rightarrow 0,0}(\Omega) + \sum_{\bar{J}=1}^{\infty} \sum_{\J_z=-\J}^{+\J} \sum_{L = \J-1, \J, \J+1} \alpha_{\bar{M}_{(s)} M_{(s)},\J,\J_z}j_L(k_{(s)}r)\f{Z}_{L\rightarrow \J,\J_z}(\Omega) \\
	&+  \frac{\chi_{\bar{M}_{(s)} M_{(s)}, 1\rightarrow 0, 0}}{r} \f{Z}_{1\rightarrow 0,0}(\Omega) + \sum_{\bar{J}=1}^{\infty} \sum_{\J_z=-\J}^{+\J} \sum_{L = \J-1, \J, \J+1} \frac{\chi_{\bar{M}_{(s)} M_{(s)}, L\rightarrow \J, \J_z}}{r} \f{Z}_{L\rightarrow \J, \J_z}(\Omega)
\end{align}
where $Y_{\J, \J_z}(\Omega)$ are the spherical harmonics and a detailed definition of $\f{Z}_{L\rightarrow \J, \J_z}(\Omega)$ can be found in our previous work \cite{Bicudo:2019ymo}. $\alpha_{\bar{M}_{(s)} M_{(s)},\J,\J_z}$ denotes the expansion coefficient for an incoming $\bar{M}_{(s)} M_{(s)}$-wave with angular momentum $\J$ and $j_L(k_{(s)} r)$ denote spherical Bessel functions.
\color{black}
The Schrödinger equation \eqref{eqn:schroedinger_equation} can then be projected in a straightforward way to definite $\J$,
\begin{align}
& \left(\frac{1}{2} \mu^{-1} \bigg(\partial_r^2 + \frac{1}{r^2} L_{\J}^2\bigg) + V_{\J}(r) +
\left(\begin{array}{ccccc}
E_{\text{threshold}} & 0 & 0 & 0 & 0 \\
0 & 2m_M & 0 & 0 & 0 \\
0 & 0 & 2m_M & 0 & 0 \\
0 & 0 & 0 & 2m_{M_s} & 0 \\
0 & 0 & 0 & 0 & 2m_{M_s} \\
\end{array}\right) 
- E\right)
\left(\begin{array}{c} u_{\J}(r) \\ \chi_{\bar{M} M,\J-1 \rightarrow \J}(r) \\ \chi_{\bar{M} M,\J+1 \rightarrow \J}(r) \\ \chi_{\bar{M}_s M_s,\J-1 \rightarrow \J}(r) \\ \chi_{\bar{M}_s M_s,\J+1 \rightarrow \J}(r) \end{array}\right) = 
\nonumber \\
& \quad \quad = \left(\begin{array}{c} V_{\textrm{mix}}(r) \\ 0 \\ 0 \\ 0 \\ 0 \end{array}\right) \bigg(\alpha_{\bar{M} M,\J-1} {\J \over 2 \J+1} r j_{\J-1}(kr) + \alpha_{\bar{M} M,\J+1} {\J+1 \over 2 \J+1} r j_{\J+1}(kr)
\nonumber \\
& \quad \quad \quad \quad \quad \quad \quad \quad \quad \quad + \alpha_{\bar{M}_s M_s,\J-1} {\J \over 2 \J+1} \frac{ r j_{\J-1}(k_s r)}{\sqrt{2}} + \alpha_{\bar{M}_s M_s,\J+1} {\J+1 \over 2 \J+1} \frac{r j_{\J+1}(k_s r)}{\sqrt{2}}\bigg)
\label{eqn:final_equation}
\end{align}
with $\mu^{-1} = \textrm{diag}(1/\mu_Q, 1/\mu_M, 1/\mu_M, 1/\mu_{M_s}, 1/\mu_{M_s})$, \\ $L_\J^2 = \textrm{diag}(\J(\J+1), (\J-1)\J, (\J+1)(\J+2), (\J-1)\J, (\J+1)(\J+2))$ and
\begin{eqnarray}
	V_\J(r) &= 
	\left(\begin{array}{ccccc}
	V_{\bar{Q} Q} & \sqrt{ {\J \over 2 \J+1} } V_\textrm{mix}  & \sqrt{{\J+1 \over 2\J+1}} V_\textrm{mix} & {1 \over \sqrt{2}} \sqrt{{\J \over 2 \J+1}} V_\textrm{mix}  & {1 \over \sqrt{2}} \sqrt{{\J+1 \over 2\J+1}}V_\textrm{mix} \\
	\sqrt{{\J \over 2 \J+1}}  V_\textrm{mix}  & 0 & 0 & 0 & 0 \\
	\sqrt{ {\J+1 \over 2\J+1}}  V_\textrm{mix} & 0 & 0 & 0 & 0 \\
	{1 \over \sqrt{2}} \sqrt{{\J \over 2 \J+1}} V_\textrm{mix}  & 0 & 0 & 0 & 0 \\
	{1 \over \sqrt{2}} \sqrt{{\J+1 \over 2\J+1}} V_\textrm{mix} & 0 & 0 & 0 & 0 \\
	\end{array}\right) .
\end{eqnarray}
Note that, equation \eqref{eqn:final_equation} is degenerate with respect to $\J_z$ thus we dropped the index.
The confining $\bar{Q} Q$ channel is represented by the radial wave function $u_{\J}(r)$ with the following boundary conditions:
\begin{itemize}
\item For $r \rightarrow 0$:
\begin{align}
u_{\J}(r) \propto r^{\J+1} .
\end{align}
\item For $r \rightarrow \infty$:
\begin{align}
u_{\J}(r) = 0 .
\end{align}
\end{itemize}
The incident wave becomes a superposition of spherical waves represented by Bessel functions $j_{L_{\textrm{in}}}$. These include $\bar M M$ waves with $L_{\textrm{in}} = \J - 1$ as well as $L_{\textrm{in}} = \J + 1$ and $\bar M_s M_s$ waves with $L_{\textrm{in}} = \J - 1$ as well as $L_{\textrm{in}} = \J + 1$, where $L_{\textrm{in}}$ denotes orbital angular momentum. Spherical waves with $L_{\textrm{in}} = \J$ are excluded, because of parity. For example, an incident $\bar M M$ wave with $L_{\textrm{in}} = \J - 1$ translates to $\vec{\alpha} = (\alpha_{\bar{M} M,\J-1}, \alpha_{\bar{M} M,\J+1}, \alpha_{\bar{M}_s M_s,\J-1}, \alpha_{\bar{M}_s M_s,\J+1}) = (1,0,0,0)$. The momenta of these waves, $k$ and $k_s$, are related to the energy $E$ via 
\begin{align}
	E = 2 m_M + \frac{k^2}{2 \mu_M} \quad , \quad E = 2 m_{M_s} + \frac{k_s^2}{2 \mu_{M_s}} .
\end{align}
The emergent wave is described by the four radial wave functions $\chi_{\bar{M}_{(s)} M_{(s)},L_{\textrm{out}} \rightarrow \J}(r)$ with $\bar{M}_{(s)} M_{(s)} \in \{ \bar{M} M , \bar{M}_s M_s \}$ and $L_{\textrm{out}} \in \{ \J-1 , \J+1 \}$, where the subscript indicates the meson content and the coupling of orbital angular momentum $L_{\textrm{out}}$ and light quark spin $S_q = 1$ to total angular momentum $\J$. The boundary conditions for the emergent wave can be formulated as follows:
\begin{itemize}
\item For $r \rightarrow 0$:
\begin{align}
\chi_{\bar{M}_{(s)} M_{(s)},L_{\textrm{out}} \rightarrow \J} \propto r^{L_{\textrm{out}}+1} .
\end{align}
\item For $r \rightarrow \infty$:
\begin{align}
	\label{EQN_t1} & \chi_{\bar{M} M,L_{\textrm{out}} \rightarrow \J} = i t_{\bar{M}_{(s)} M_{(s)}, L_{\textrm{in}}; \bar M M, L_{\textrm{out}}} r h^{(1)}_{L_{\textrm{out}}}(k r) \\
	\label{EQN_t2} & \chi_{\bar{M}_s M_s,L_{\textrm{out}} \rightarrow \J} =  i t_{\bar{M}_{(s)} M_{(s)}, L_{\textrm{in}}; \bar M_s M_s, L_{\textrm{out}}} r h^{(1)}_{L_{\textrm{out}}}(k_s r) ,
\end{align}
where
\begin{itemize}
\item $\bar{M}_{(s)} M_{(s)} \equiv \bar{M} M$ for $\vec{\alpha} = (1,0,0,0)$ and $\vec{\alpha} = (0,1,0,0)$, i.e.\ an incident $\bar M M$ wave with $L_{\textrm{in}} = \J - 1$ and $L_{\textrm{in}} = \J + 1$, respectively,
\item $\bar{M}_{(s)} M_{(s)} \equiv \bar{M}_s M_s$ for $\vec{\alpha} = (0,0,1,0)$ and $\vec{\alpha} = (0,0,0,1)$, i.e.\ an incident $\bar M_s M_s$ wave with $L_{\textrm{in}} = \J - 1$ and $L_{\textrm{in}} = \J + 1$, respectively.
\end{itemize}
\end{itemize}
These boundary conditions define 16 quantities, $t_{\bar{M}_{(s)} M_{(s)}, L_{\textrm{in}}; \bar M_{(s)} M_{(s)}, L_{\textrm{out}}}$, which represent scattering amplitudes and can be combined to the $4\times 4$ $\mbox{T}$ matrix
\begin{align}
	\mbox{T}_\J = 
	\left(\begin{array}{cccc}
	t_{\bar{M} M, \J-1; \bar{M} M, \J-1} & t_{\bar{M} M, \J+1; \bar{M} M, \J-1}  & t_{\bar{M}_s M_s, \J-1; \bar{M} M, \J-1} & t_{\bar{M}_s M_s, \J+1; \bar{M} M, \J-1} \\
	t_{\bar{M} M, \J-1; \bar{M} M, \J+1} & t_{\bar{M} M, \J+1; \bar{M} M, \J+1}  & t_{\bar{M}_s M_s, \J-1; \bar{M} M, \J+1} & t_{\bar{M}_s M_s, \J+1; \bar{M} M, \J+1} \\
	t_{\bar{M} M, \J-1; \bar{M}_s M_s, \J-1} & t_{\bar{M} M, \J+1; \bar{M}_s M_s, \J-1}  & t_{\bar{M}_s M_s, \J-1; \bar{M}_s M_s, \J-1} & t_{\bar{M}_s M_s, \J+1; \bar{M}_s M_s, \J-1} \\
	t_{\bar{M} M, \J-1; \bar{M}_s M_s, \J+1} & t_{\bar{M} M, \J+1; \bar{M}_s M_s, \J+1}  & t_{\bar{M}_s M_s, \J-1; \bar{M}_s M_s, \J+1} & t_{\bar{M}_s M_s, \J+1; \bar{M}_s M_s, \J+1} \\
	\end{array}\right) .
	\label{eqn:t_matrix_4x4}
\end{align}
The $\mbox{T}$ matrix is related to the $\mbox{S}$ matrix in the usual way,
\begin{align}
\mbox{S}_\J = 1 + 2 i \mbox{T}_\J .
\end{align}

For $\J = 0$ one has to discard the contributions to the incident wave with $L_{\textrm{in}} = \J-1$ and to the emergent wave with $L_{\textrm{out}} = \J-1$. The Schr\"odinger equation \eqref{eqn:final_equation} is then reduced from five to three channels and $\mbox{T}_\J$ from a $4 \times 4$ to a $2 \times 2$ matrix. This $\J = 0$ equation is extensively discussed in Ref.\ \cite{Bicudo:2020qhp}.

In analogy to a single channel scattering problem, where the phase shift $\delta$ is defined via $1 + 2 i \mbox{T} = \mbox{S} = \exp(2 i \delta)$, one can define the eigenphase sum \cite{PhysRevA.19.920,Ashton:1983,Hagino:2019tbn} for multi-channel scattering via
\begin{align}
\textrm{det}(\mbox{S}_\J) = \exp(2 i \delta_\J) .
\label{eqn:eigen_delta}
\end{align}
The eigenphase sum $\delta_\J$ is identical to a sum of phase shifts, where each phase shift corresponds to one of the eigenvalues of the $\mbox{S}$ matrix.

\section{\label{SEC003}Numerical results}

\subsection{Numerical methods to solve the coupled channel Schrödinger equation and to determine the poles of the $\mbox{T}$ matrix}

In Section~\ref{SEC390} we defined the entries of the $\mbox{T}$ matrix (\ref{eqn:t_matrix_4x4}) as the a priori unknown coefficients \\ $t_{\bar{M}_{(s)} M_{(s)}, L_{\textrm{in}}; \bar M_{(s)} M_{(s)}, L_{\textrm{out}}}$ appearing in the $r \rightarrow \infty$ boundary conditions (\ref{EQN_t1}) and (\ref{EQN_t2}). To determine these coefficients, one has to solve the coupled channel Schrödinger equation (\ref{eqn:final_equation}). To cross check our results, we used two rather different numerical methods. The first method corresponds to discretizing the radial coordinate by a uniform grid and solving the resulting system of linear equations using methods from standard textbooks (for details see Ref.\ \cite{Bicudo:2019ymo}). The second method corresponds to using an ordinary 4th order Runge-Kutta algorithm.

To find the poles of $\mbox{T}_\J$ in the complex energy plane, chracterized by at least one of its eigenvalues approaching infinity, we applied the Newton-Raphson method to find the roots of $1/\textrm{det}(\mbox{T}_\J)$.

\subsection{\label{SEC433}Input parameters and error analysis} 

In the following we present results for heavy $b$ quarks, i.e.\ $Q \equiv b$. We use $m_Q = 4.977 \,\text{GeV}$ from quark models \cite{PhysRevD.32.189} and the spin-averaged mass of the $B$ meson and the $B^*$ meson, i.e.\ $m_M = (m_B + 3 m_{B^*})/4 = 5.313 \, \text{GeV}$, as well as of the $B_s$ meson and the $B_s^*$ meson, i.e.\ $m_{M_s} = (m_{B_s} + 3 m_{B_s^*})/4 = 5.403 \, \text{GeV}$. The lattice data from Ref.\ \cite{Bali:2005fu} we are using to determine the parameters of the potential parameterizations (see Section~\ref{SEC488}) was generated with a light quark mass slightly below the physical strange quark mass. This is reflected by $E_{\text{threshold}} = 10.790 \, \text{GeV}$, which is much closer to the spin-averaged $B_s^{(*)} B_s^{(*)}$ threshold than to the spin-averaged $B^{(*)} B^{(*)}$ threshold.

The uncertainties of the lattice data provided in Ref.\ \cite{Bali:2005fu} are propagated via resampling. We generated 1000 statistically independent samples and repeated our computations on each of them. For our results we quote asymmetric errors, defined by the 16th and 84th percentile. 

\subsection{\label{SEC852}Eigenphase sums and poles of the $\mbox{T}$ matrix}

We have computed the eigenphase sum $\delta_\J$ defined in Eq.\ \eqref{eqn:eigen_delta} for $\J = 0,1,2,3$ as function of the energy above the spin-averaged $B^{(*)} B^{(*)}$ threshold at $10.627 \, \text{GeV}$. We show the respective plots in Fig.\ \ref{fig:scattering_phase}. For rather stable resonances, which are clearly separated in their energies, such plots show pronounced steps of order $\pi$. The locations of these steps then correspond to the resonance masses and the slopes are inversely proportional to the associated decay widths. In our case, however, where some resonances have large decay widths and their energy levels are close, it is hard to identify them in clear and unique way.

\begin{figure*}
	\includegraphics[width=0.48\textwidth]{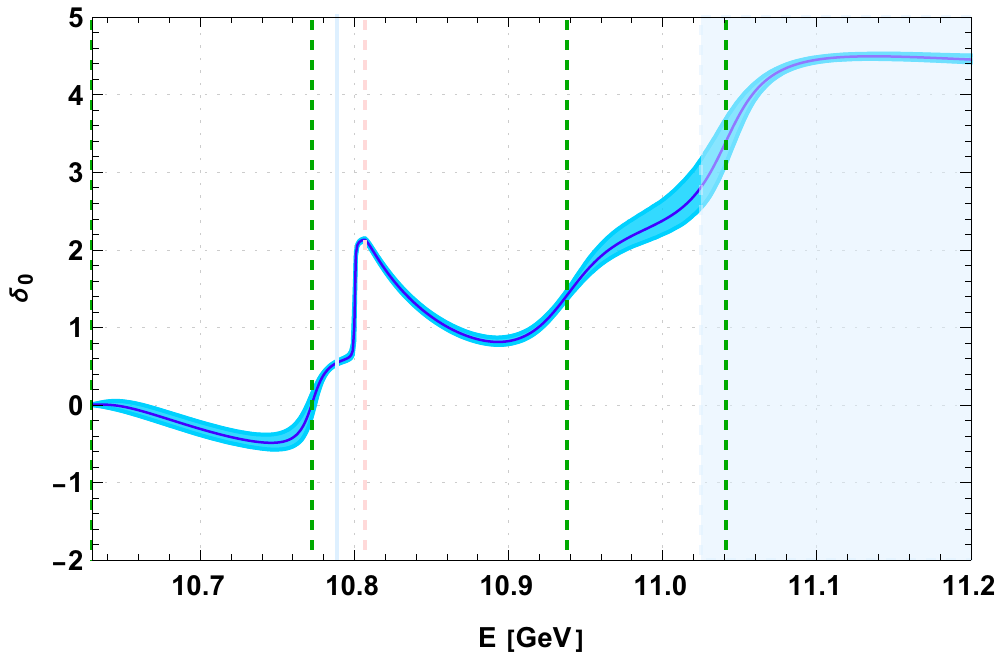}
	\includegraphics[width=0.48\textwidth]{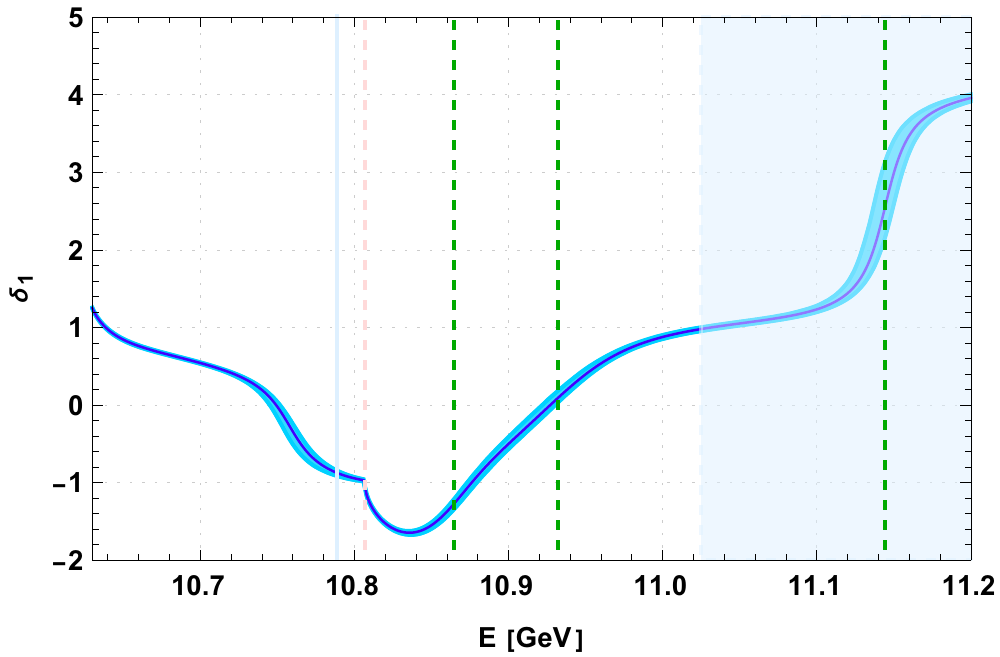}
	\includegraphics[width=0.48\textwidth]{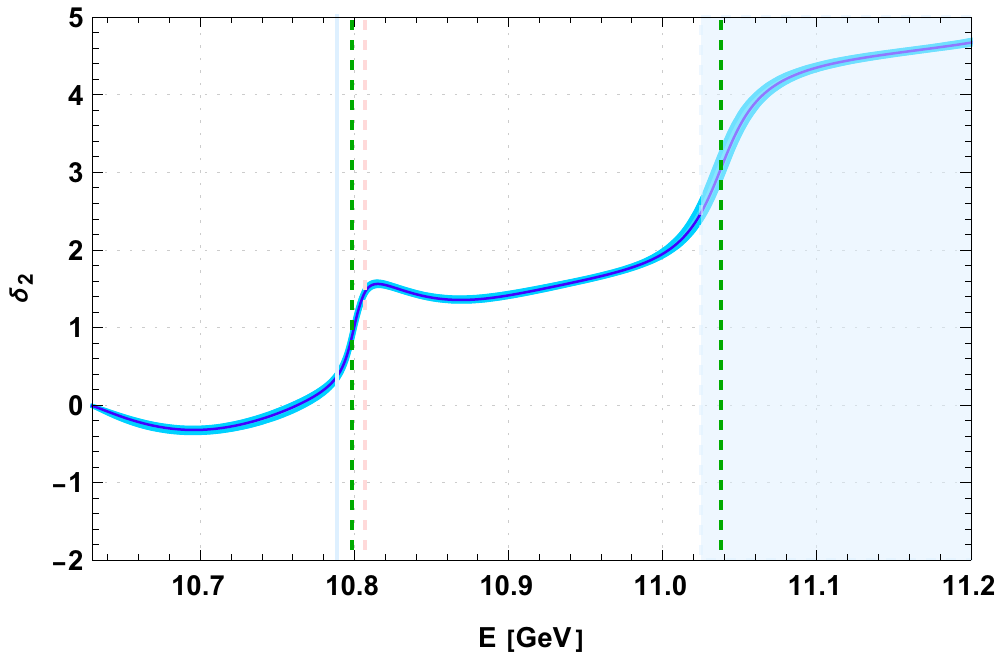}
	\includegraphics[width=0.48\textwidth]{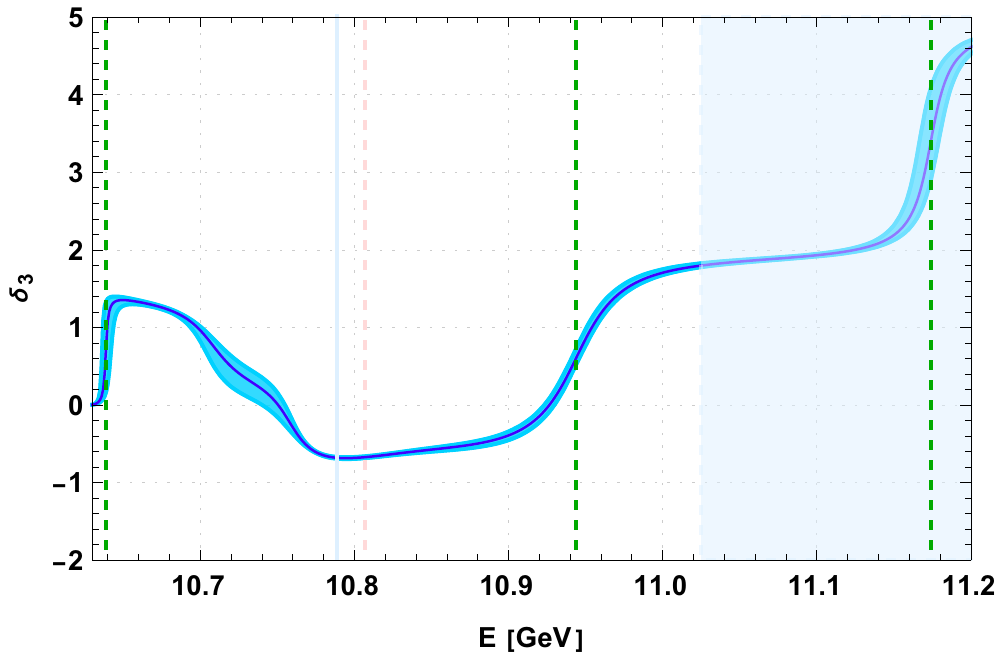}
	\caption{\label{fig:scattering_phase}Eigenphase sum $\delta_\J$ as function of the energy $E$. Narrow resonances are indicated by pronounced steps. We also show the real parts of the positions of the poles of $\mbox{T}_\J$ using green dotted lines (the line at $\approx 10.798 \, \text{GeV}$ in the lower left plot does not correspond to the real part of a pole, but to a fit to the eigenphase sum; see the technical discussion in Section~\ref{SEC599}). A pale red dotted line indicates the spin averaged $B_s^{(*)}B_s^{(*)}$ threshold, the light blue solid line $E_\text{threshold}$. Results for energies inside the light-blue shaded region, i.e.\ above $\approx 11.025 \, \text{GeV}$, should not be trusted, since we neglect decay channels containing a negative and a positive parity heavy-light meson. }
\end{figure*}

To define and to compute masses and decay widths of bottomonium resonances in a clearer and more definite way, we have analytically continued our scattering problem to the complex energy plane. Then we have determined the poles of the $\mbox{T}$ matrix \eqref{eqn:t_matrix_4x4} in the complex energy plane numerically. The positions of these poles can be related to masses and decay widths according to 
\begin{align}
m = \textrm{Re}(E_\text{pole}) \quad , \quad \Gamma = -2 \textrm{Im}(E_\text{pole}) ,
\end{align}
where $E_\text{pole}$ denotes the complex pole energy.

In Fig.\ \ref{fig:poles} we show all poles of the $\mbox{T}$ matrix \eqref{eqn:t_matrix_4x4} for $\J = 0,1,2,3$ up to $11.2 \, \text{GeV}$. Colored point clouds represent 1000 independent computations with resampled lattice data from Ref.\ \cite{Bali:2005fu}. These point clouds are used to determine statistical errors, which is straightforward, because there are clear gaps between the point clouds. Bound states correspond to poles located on the real axis below the $B^{(*)} B^{(*)}$ threshold, while resonances correspond to poles above this threshold with a non-vanishing negative imaginary part. The pole positions and their statistical errors are indicated by the black crosses. The $B^{(*)} B^{(*)}$ threshold at $10.627 \, \text{GeV}$ and the $B_s^{(*)} B_s^{(*)}$ threshold at $10.807 \, \text{GeV}$ are indicated by pale red dotted lines. Our results can only be trusted up to the threshold of one negative parity heavy-light meson and one positive parity heavy-light meson at around $11.025 \, \text{GeV}$, because the corresponding decay channel is not included in our Schrödinger equation (\ref{eqn:schroedinger_equation}). The region above this threshold is shaded in light-blue.

We summarize these theoretical predictions of bottomonium masses and decay widths in Table~\ref{tab:polepositions}, together with available experimental results. Additionally, we show the same data in a graphical way in Fig.\ \ref{fig:spectrum}.

\begin{figure*}
	\includegraphics[width=0.48\textwidth]{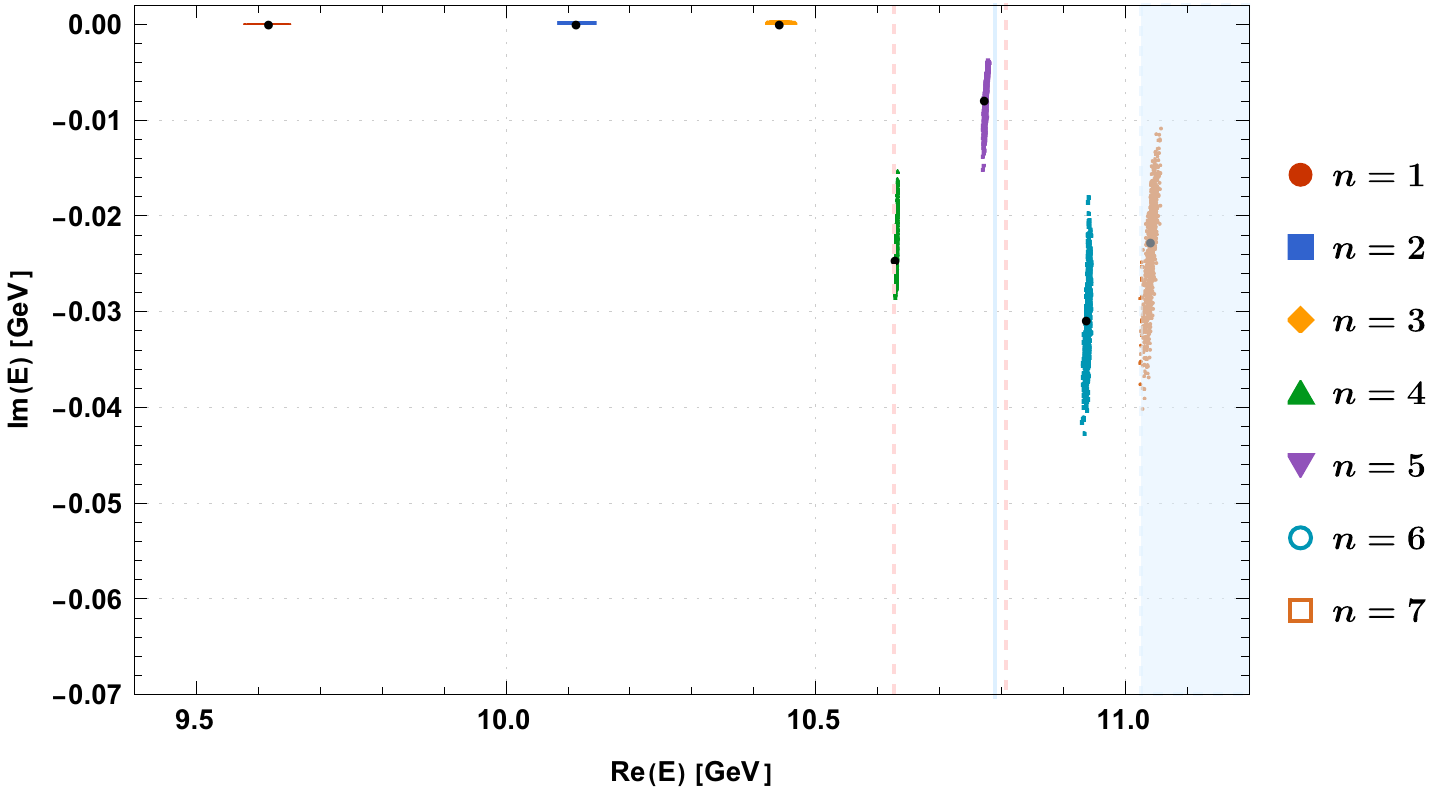}
	\includegraphics[width=0.48\textwidth]{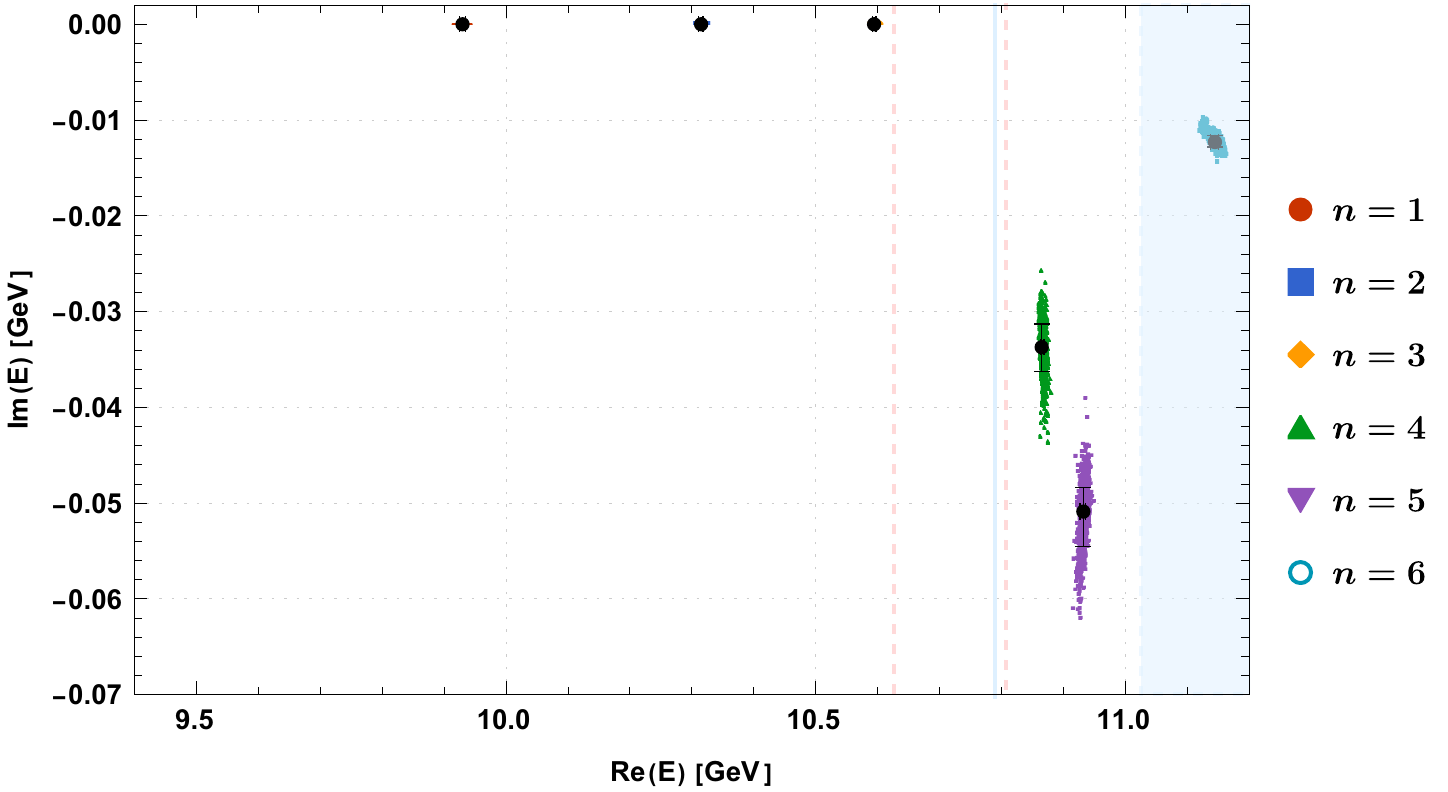}
	\includegraphics[width=0.48\textwidth]{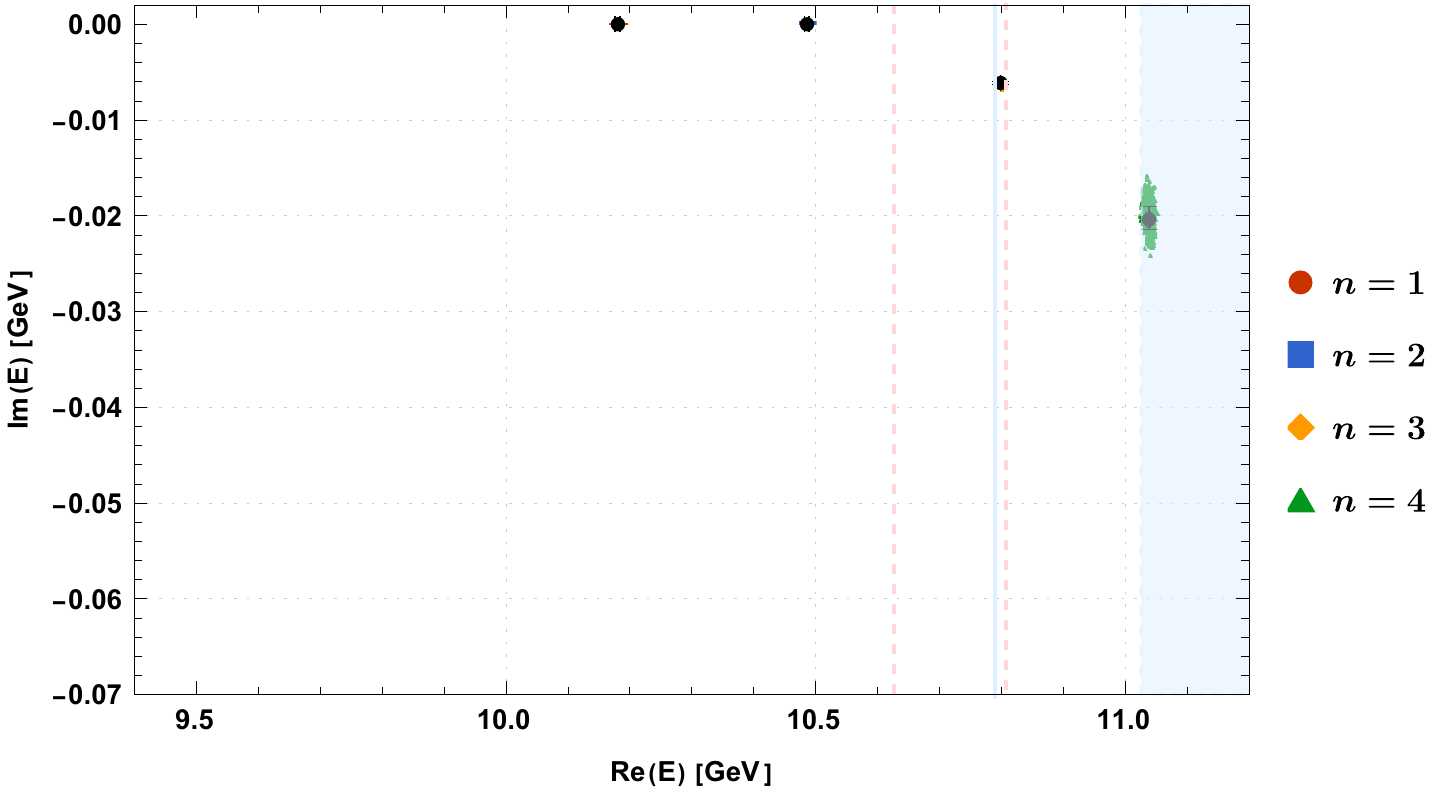}
	\includegraphics[width=0.48\textwidth]{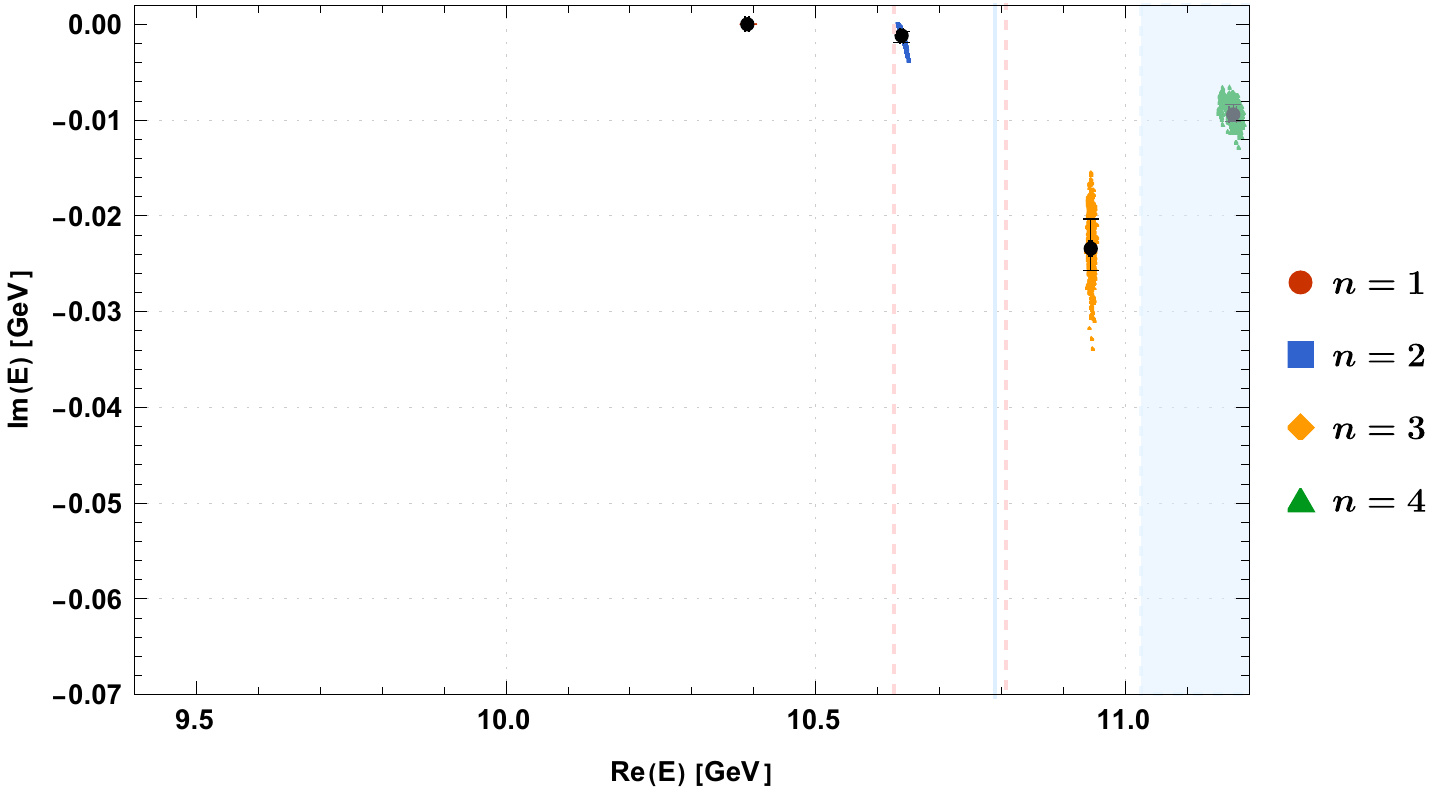}
	\caption{\label{fig:poles}Positions of the poles of the $\mbox{T}$ matrix (\ref{eqn:t_matrix_4x4}) in the complex energy plane for $\J=0$ (top left), $\J=1$ (top right), $\J=2$ (bottom left) and $\J=3$ (bottom right) representing bound states and resonances up to $11.2 \, \text{GeV}$. Colored point clouds represent 1000 independent computations with resampled lattice data, while black dots correspond to the mean values and error bars. The pale red dotted lines indicate the spin averaged $B^{(*)}B^{(*)}$ and $B_s^{(*)}B_s^{(*)}$ thresholds, the light blue solid line $E_\text{threshold}$. Results for energies inside the light-blue shaded region, i.e.\ above $\approx 11.025 \, \text{GeV}$, should not be trusted, since we neglect decay channels containing a negative and a positive parity heavy-light meson.}
\end{figure*}

\begin{table*}
	\includegraphics[width=0.7\textwidth]{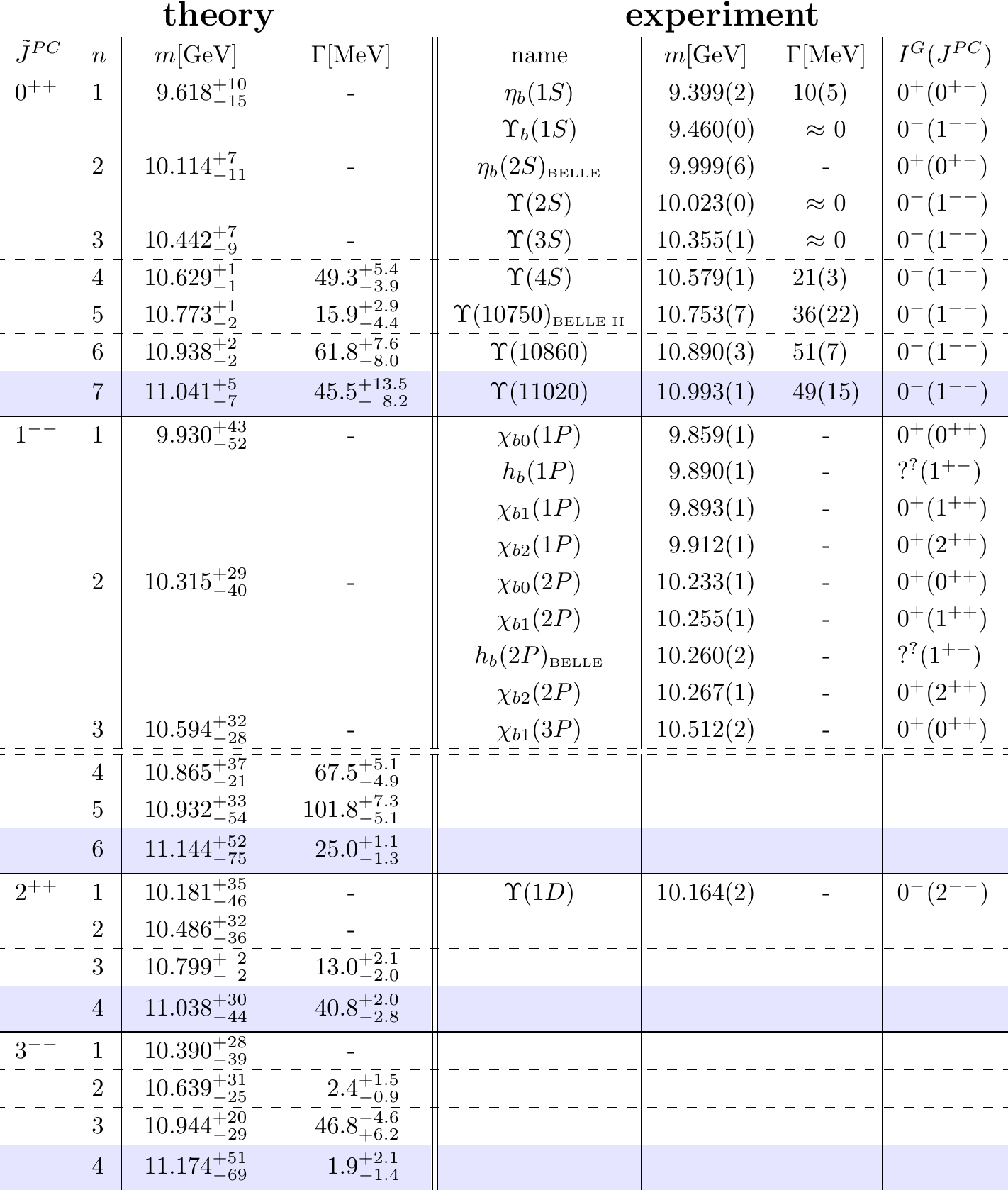}
	\caption{\label{tab:polepositions}Masses and decay widths of $I = 0$ bottomonium with $\J^{PC}=0^{++},1^{--},2^{++},3^{--}$ obtained from the poles of the $\mbox{T}$ matrix (\ref{eqn:t_matrix_4x4}), where errors are purely statistical. An exception is the $n = 3$ resonance for $\J = 2$, which was extracted by a fit to the eigenphase sum $\delta_2$ (see the technical discussion in Section~\ref{SEC599}). The spin-averaged $B^{(*)}B^{(*)}$ and $B_s^{(*)}B_s^{(*)}$ thresholds are indicated by dashed lines. Results for energies above the threshold of one negative and another positive heavy-light meson at $\approx 11.025 \, \text{GeV}$ are marked by a light-blue background and should not be trusted, since we neglect the corresponding decay channels. (Results for $\J^{PC}=0^{++}$ were already presented in Ref.\ \cite{Bicudo:2020qhp}.) }
\end{table*}

\begin{figure*}
	\includegraphics[width=\textwidth]{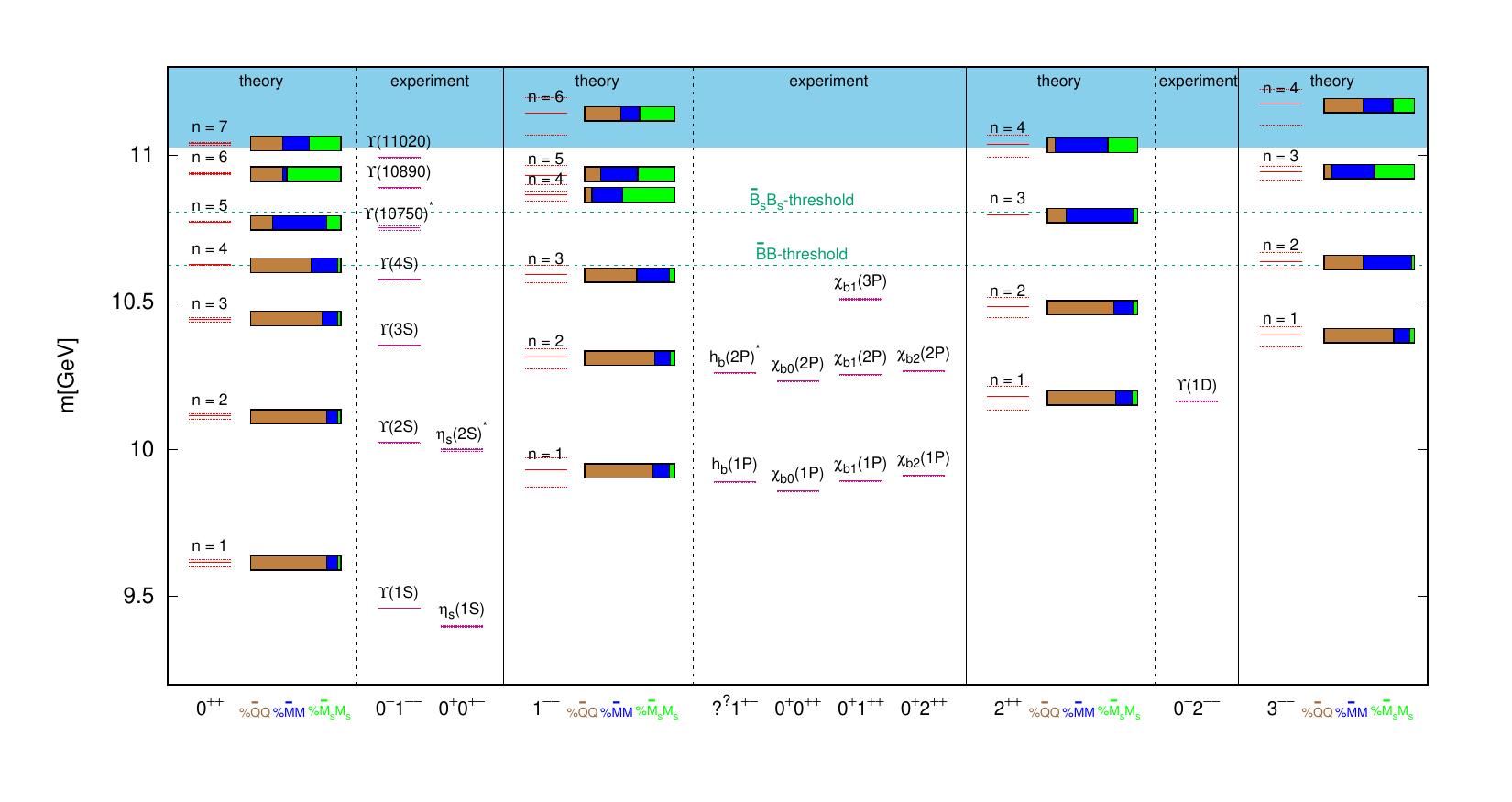}
	\caption{\label{fig:spectrum}Graphical summary of theoretical predictions and experimental results for masses of $I = 0$ bottomonium with $\J^{PC}=0^{++},1^{--},2^{++},3^{--}$. We also show the quarkonium and meson-meson composition as discussed in Section~\ref{sec:percentages}: $\% \bar Q Q$ in orange, $(\% \bar M M)_{\J-1} + (\% \bar M M)_{\J+1}$ in blue and $(\% \bar M_s M_s)_{\J-1} + (\% \bar M_s M_s)_{\J+1}$ in green. }
	\end{figure*}

\subsubsection{\label{SEC599}Technical aspects of pole finding}

In addition to the physical poles of the $\mbox{T}$ matrix, which correspond to bottomonium bound states and resonances, there are also unphysical poles, which are caused by numerical inaccuracies. Such unphysical poles can be identified, by varying numerical parameters (e.g.\ the spacing of the grid, the Runge-Kutta step size or the large, but finite value of $r$ replacing $r \rightarrow \infty$ in the boundary conditions (\ref{EQN_t1}) and (\ref{EQN_t2})). If a pole is unstable with respect to these parameters, it is clearly an unphysical pole.

In cases, where a physical pole and unphysical poles are close, pole finding might become a delicate task. To verify that we did not miss any of the physical poles with our pole finding algorithm, we have also determined the poles after removing the $\bar B^{(*)}_s B^{(*)}_s$ decay channels. In general, the pole positions in the 3-flavor case are similar to those in the 2-flavor case and consistent with the steps observed in the eigenphase sums shown in Fig.\ \ref{fig:scattering_phase}. An exception is the $n = 3$ resonance in the $\J = 2$ sector, which or pole finding algorithm was unable to identify in the 3-flavor case. It seems to be masked by unphysical poles. Thus, to determine the corresponding resonance parameters, we have used the eigenphase sum $\delta_2$. We have performed a 4-parameter fit with $\alpha + \beta \text{arctan}((2 / \Gamma) (E-m))$ to the data points for $\delta_2$ in the region of the step at $\approx 10.798 \, \text{GeV}$ (see Fig.\ \ref{fig:polefromeigenphase}). In this way we have obtained the resonance mass $m = 10.799_{-2}^{+2} \, \text{GeV}$ and the decay width $\Gamma = 13.0_{-2.0}^{+1.8} \, \text{MeV}$, where error bars are determined by resampling lattice data from Ref.\ \cite{Bali:2005fu} and by varying the fit range. As a cross check we have also determined the parameters of the $n = 4$ resonance in the $\J = 2$ sector in the same way and find $m = 11.038_{-3}^{+5} \, \text{GeV}$ and $\Gamma = 42.29_{-1.2}^{+2.0} \, \text{MeV}$. These results are fully consistent with those obtained from our pole search (cf.\ Table~\ref{tab:polepositions}).

\begin{figure}
	\includegraphics[width=0.47\textwidth]{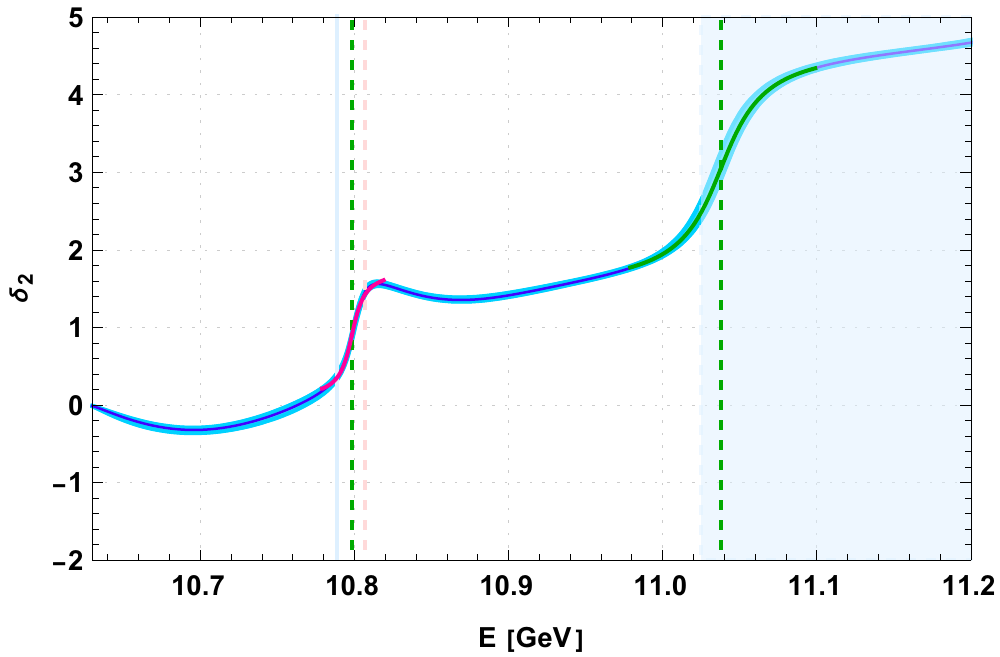}
	\caption{\label{fig:polefromeigenphase}Fits of $\alpha + \beta \text{arctan}((2 / \Gamma) (E-m))$ to the data points for the eigenphase sum $\delta_2$ in the region of the steps at $\approx 10.798 \, \text{GeV}$ and $\approx 11.038 \, \text{GeV}$. }
\end{figure}

\subsubsection{Comparison to experimental results}

For the predicted low-lying states it is straightforward to assign experimental counterparts.
\begin{itemize}
\item The $\J = 0$ states with $n=1,2,3,4$ correspond to $\eta_b(1S) \equiv \Upsilon(1S)$, to $\Upsilon(2S)$, to $\Upsilon(3S)$ and to $\Upsilon(4S)$.

\item The $\J = 1$ states with $n=1,2,3$ correspond to $h_b(1P) \equiv \chi_{b0}(1P) \equiv \chi_{b1}(1P) \equiv \chi_{b2}(1P)$, to \\ $h_b(2P) \equiv \chi_{b0}(2P) \equiv \chi_{b1}(2P) \equiv \chi_{b2}(2P)$ and to $\chi_{b1}(3P)$.

\item The $\J = 2$ state with $n=1$ corresponds to $\Upsilon(1D)$.	
\end{itemize}
Our masses exhibit a pattern, which is quite similar to that found in experiments. The largest discrepancies are observed for the lowest states, most prominently for the ground state $\eta_b(1S) \equiv \Upsilon(1S)$. This, however, does not indicate particular problems with these states, but is rather a consequence of choosing $E_\text{threshold} \approx 2 m_{M_s}$ as reference point to introduce the energy scale. Compared to this reference point the errors of our predicted states are of order $10 \%$. To a large part these errors can be compensated by a global multiplicative factor. A possible reason for this error might be the scale setting in the lattice QCD computation of Ref.\ \cite{Bali:2005fu}, which is based on defining $r_0 = 0.5 \, \text{fm}$, while more recent lattice investigations indicate a smaller value for $r_0$ \cite{Sommer:2014mea}. Using e.g.\ $r_0 = 0.45 \, \text{fm}$ instead of $r_0 = 0.5 \, \text{fm}$ reduces the error for $\eta_b(1S) \equiv \Upsilon(1S)$ by around $50\%$. Moreover, our potentials were determined from lattice QCD data obtained at a single coarse lattice spacing and with only two light quark flavors close to the mass of the physical strange quark. Thus, for precise quantitative predictions a more accurate lattice QCD computation of the relevant potentials and mixing angles will be necessary (see also the outlook in Section~\ref{SEC004}).

The $\J = 0$ resonance with $n = 5$ has a mass close to the experimental result for $\Upsilon(10753)$, which was recently reported by Belle \cite{Abdesselam:2019gth}. In a previous publication we investigated the structure of this state within the same setup and found that it is meson-meson dominated with just a small quark-antiquark component \cite{Bicudo:2020qhp} (see also Section~\ref{sec:percentages}, in particular Table~\ref{tab:percentages}). Thus, since it is not an ordinary quarkonium state and the heavy quark spin can be $1^{--}$, it can be classified as a $Y$ type crypto-exotic state. Note that we found another resonance in that energy region with quantum numbers $\J = 2$ and $n = 3$. This state is, however, farther away from the experimental result for $\Upsilon(10753)$ ($\approx 46 \, \text{MeV}$ difference for $\J = 2$, $n = 3$ compared to $\approx 20 \, \text{MeV}$ difference for $\J = 0$, $n = 5$) and, thus, an identification with $\Upsilon(10753)$ seems less likely. On the other hand, since our results exhibit certain systematic errors, as discussed in the previous paragraph, we are not in a position to fully exclude such an identification.

The resonances $\Upsilon(10860)$ and $\Upsilon(11020)$ are typically interpreted as $\Upsilon(5S)$ and $\Upsilon(6S)$. However, from the experimental perspective they could as well correspond to $D$ wave states. The $\J = 0$ $S$ wave resonance with $n = 6$ is rather close to the mass of $\Upsilon(10860)$, whereas there is no $\J = 2$ $D$ wave resonance in that energy region. Thus, our results support the interpretation of $\Upsilon(10860)$ as $\Upsilon(5S)$.
Concerning $\Upsilon(11020)$, the $\J = 0$ resonance with $n = 7$ and the $\J = 2$ resonance with $n = 4$ have almost the same mass and are both close to the mass of $\Upsilon(11020)$. Moreover, that mass is already close to the threshold of a negative parity $B$ or $B^\ast$ and a positive parity $B_0^\ast$ or $B_1^\ast$ meson, a channel we have not yet included in our approach. Thus, we cannot decide, whether the $\Upsilon(11020)$ is indeed an $S$ wave or rather a $D$ wave state.

\subsection{\label{sec:percentages}Quarkonium and meson-meson composition}

Using techniques developed in Ref.\ \cite{Bicudo:2020qhp} we also study the structure and quark content of bound states and resonances, to clarify, whether they are conventional quarkonia or there are sizable $\bar Q Q \bar q q$ four-quark components. To this end, we compute for each state the percentages of quarkonium with $L = \J$ and of $\bar{M} M$ and $\bar{M}_s M_s$ meson-meson pairs with $L_{\text{out}} = \J-1, \J+1$,
\begin{align}
	\% \bar Q Q &= \frac{Q}{Q + M_{\J-1}+ M_{\J+1}+ M_{s,\J-1}+ M_{s,\J+1}} \\
	(\% \bar M M)_{L_{\text{out}}} &= \frac{M_{L_{\text{out}}}}{Q + M_{\J-1}+ M_{\J+1}+ M_{s,\J-1}+ M_{s,\J+1}} \\
	(\% \bar M_s M_s)_{L_{\text{out}}} &= \frac{M_{s,L_{\text{out}}}}{Q + M_{\J-1}+ M_{\J+1}+ M_{s,\J-1}+ M_{s,\J+1}} ,
\end{align}
where
\begin{align}
	Q &= \int_0^{R_{\text{max}}} \text{d}r \, \Big|u_{\J}(r)\Big|^2 \\
	M_{L_{\text{out}}} &= \int_0^{R_{\text{max}}} \text{d}r \, \Big|\chi_{\bar M M, L_{\text{out}} \rightarrow \J}(r)\Big|^2 \\
	M_{s,L_{\text{out}}} &= \int_0^{R_{\text{max}}} \text{d}r \, \Big|\chi_{\bar M_s M_s, L_{\text{out}} \rightarrow \J}(r)\Big|^2 .
\end{align}
$u_{\J}(r)$ , $\chi_{\bar M M, L_{\text{out}} \rightarrow \J}(r)$ and $\chi_{\bar M_s M_s, L_{\text{out}} \rightarrow \J}(r)$ are the solutions of the radial Schrödinger equation (\ref{eqn:final_equation}) for real energy $\text{Re}(E_\text{pole})$, where $E_\text{pole}$ is the position of the corresponding pole of the $\mbox{T}$ matrix in the complex energy plane. 

In case of a bound state, $Q$, $M_{L_{\text{out}}}$ and $M_{s,L_{\text{out}}}$ approach constants for $R_{\text{max}} \gtapprox 2.0 \, \text{fm}$ as indicated by Figure~\ref{fig:percentages_J0} to Figure~\ref{fig:percentages_J3}. The corresponding asymptotic values of $\% \bar Q Q$, $(\% \bar M M)_{L_{\text{out}}}$ and $(\% \bar M_s M_s)_{L_{\text{out}}}$ for bound states are collected in Table~\ref{tab:percentages}.

\begin{table*}
\includegraphics[width=\textwidth]{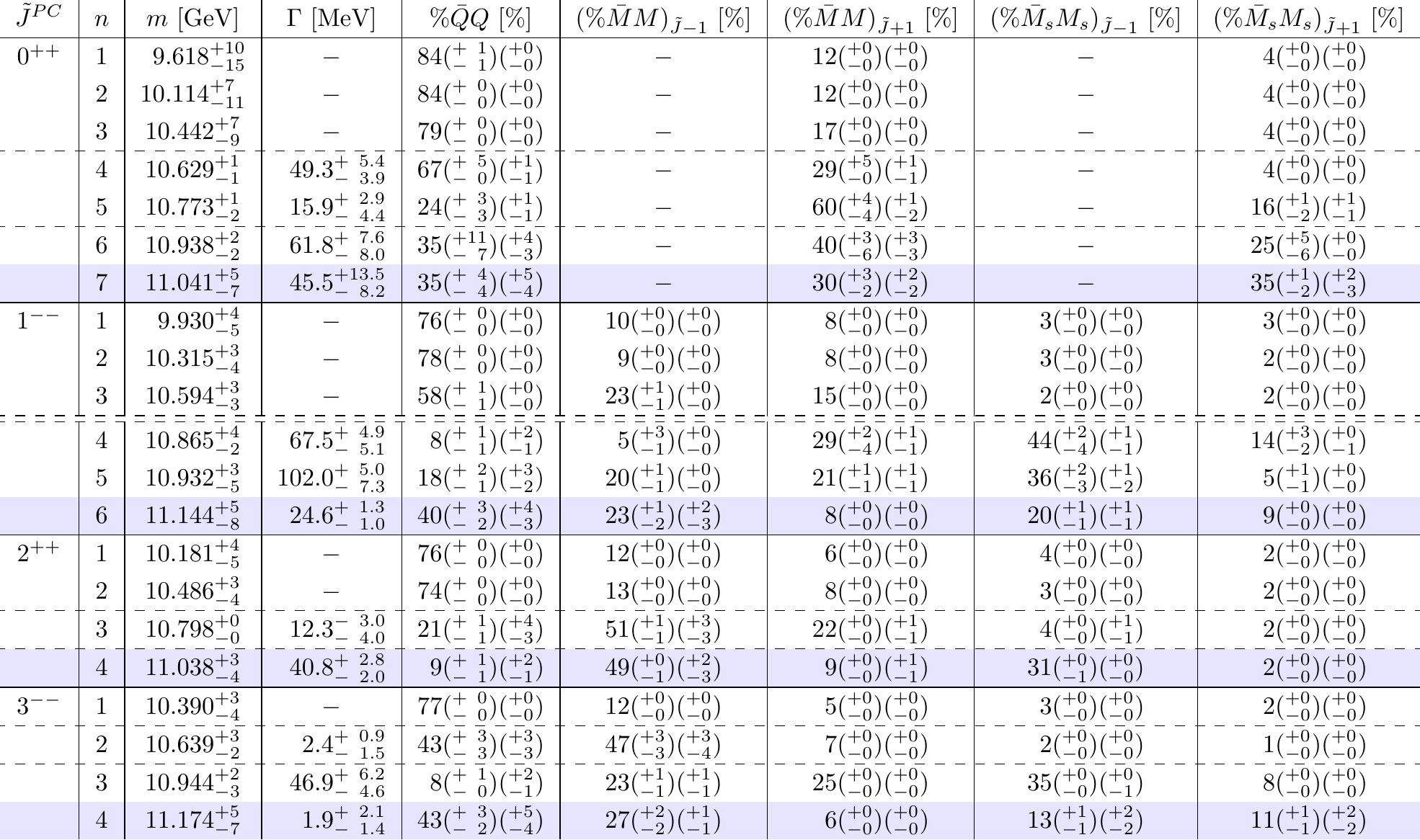}
\caption{\label{tab:percentages}Percentages of quarkonium and of meson-meson pairs for $I = 0$ bottomonium with $\J^{PC}=0^{++},1^{--},2^{++},3^{--}$. In addition to statistical errors we also provide systematic uncertainties, which are discussed in Section~\ref{sec:percentages}. The spin-averaged $B^{(*)}B^{(*)}$ and $B_s^{(*)}B_s^{(*)}$ thresholds are indicated by dashed lines. Results for energies above the threshold of one negative and another positive heavy-light meson at $\approx 11.025 \, \text{GeV}$ are marked by a light-blue background and should not be trusted, since we neglect the corresponding decay channels. (Results for $\J^{PC}=0^{++}$ were already presented in Ref.\ \cite{Bicudo:2020qhp}.)}
\end{table*}

In case of a resonance, $M_{L_{\text{out}}}$ and $M_{s,L_{\text{out}}}$ are linearly rising functions for large $R_{\text{max}}$, because in that region $\chi_{\bar M M, L_{\text{out}} \rightarrow \J}(r)$ and $\chi_{\bar M_s M_s, L_{\text{out}} \rightarrow \J}(r)$ represent emergent spherical waves. The corresponding slopes, however, are rather small and result in changes of only a few percent in $\% \bar Q Q$, $(\% \bar M M)_{L_{\text{out}}}$ and $(\% \bar M_s M_s)_{L_{\text{out}}}$ in the interval $1.8 \, \text{fm} < R_{\text{max}} < 3.0 \, \text{fm}$. At such separations $R_{\text{max}}$ the quarkonium component $u_{\J}(r)$ is already negligible and $\chi_{\bar M M, L_{\text{out}} \rightarrow \J}(r)$ and $\chi_{\bar M_s M_s, L_{\text{out}} \rightarrow \J}(r)$ are almost pure emergent spherical waves. Thus the corresponding resonance is contained inside a sphere of radius $R_{\text{max}}$. We evaluate $\% \bar Q Q$, $(\% \bar M M)_{L_{\text{out}}}$ and $(\% \bar M_s M_s)_{L_{\text{out}}}$ at the center of this interval, i.e.\ at $R_{\text{max}} = 2.4 \, \text{fm}$, but assign asymmetric systematic uncertainties \\ $|\% \bar Q Q(R_{\text{max}} = 1.8\,\text{fm}) - \% \bar Q Q(R_{\text{max}}=2.4\,\text{fm})|$ and $|\% \bar Q Q(R_{\text{max}} = 3.0 \, \text{fm}) - \% \bar Q Q(R_{\text{max}} = 2.4 \, \text{fm})|$ to $\% \bar Q Q$ and in an analogous way also to $(\% \bar M M)_{L_{\text{out}}}$ and to $(\% \bar M_s M_s)_{L_{\text{out}}}$. These results are collected in Table~\ref{tab:percentages}. Moreover, $\% \bar Q Q$, $(\% \bar M M)_{L_{\text{out}}}$ and $(\% \bar M_s M_s)_{L_{\text{out}}}$ as functions of $R_{\text{max}}$ are shown in Figure~\ref{fig:percentages_J0} to Figure~\ref{fig:percentages_J3}.

The majority of bound states have masses significantly below the $B^{(*)}B^{(*)}$ threshold with binding energies of the order of $100 \, \text{MeV}$ or larger ($\J = 0$, $n = 1, 2, 3$; $\J = 1$, $n = 1, 2$; $\J = 2$, $n = 1, 2$; $\J = 3$, $n = 1$). These states consist mostly of quarkonium, $\% \bar Q Q \approx 74\% \ldots 84\%$. The largest quarkonium components are present for the lowest states with orbital angular momentum $L = 0$, i.e.\ for $\J = 0$ and $n = 1, 2$. Because of the non-vanishing mixing angle $\theta \approx 0.35 \ldots 0.40$ for $r \ltapprox 1.0 \, \text{fm}$ obtained by a full lattice QCD computation (see Ref.\ \cite{Bali:2005fu} and our previous article \cite{Bicudo:2019ymo}, in particular Section~III), larger quarkonium percentages $\% \bar Q Q$ are excluded. The reason is that for two light quark flavors the ground state potential in the $\Sigma_g^+$ sector, which is of central importance for bound states, is a linear superposition of $1 - \sin^2(\theta) \approx 85\% \ldots 88\%$ quarkonium and of $\theta^2 \approx 12\% \ldots 15\%$ meson-meson. Thus, a very similar composition is generated dynamically for the deeply bound states by our Schrödinger equation (\ref{eqn:schroedinger_equation}), which was derived to be consistent with the lattice QCD potentials from Ref.\ \cite{Bali:2005fu} and the corresponding mixing angle.

However, there is one bound state ($\J=1$, $n = 3$) with $m \approx 10.594 \, \text{GeV}$, rather close to the $B^{(*)} B^{(*)}$ threshold at $10.627 \, \text{GeV}$, where the quarkonium component is already significantly reduced, $\% \bar Q Q \approx 58\%$. A reason for that could be that the components of the radial wave functions are significantly farther extended, up to $r \approx 1.5 \, \text{fm} \ldots 2.0 \, \text{fm}$ (see Figure~\ref{fig:percentages_J1}). Close to the string breaking distance $r_\text{sb}$ the mixing angle changes rapidly from $\theta \approx 0.35 \ldots 0.40$ to $\theta \approx \pi/2$. Thus, for $r \gtapprox r_\text{sb} \approx 1.25 \, \text{fm}$ the ground state potential in the $\Sigma_g^+$ sector corresponds almost exclusively to a $B^{(*)} B^{(*)}$ pair. Consequently, $(\% \bar M M)_{L_{\text{out}}}$ is significantly enhanced for such a spatially extended state compared to the more tightly bound states discussed in the previous paragraph. 

The resonances with $\J=0$, $n = 4$ and $\J=3$, $n=2$ are slightly above the $B^{(*)} B^{(*)}$ threshold, i.e.\ energetically they are extremely close to bound states. In both cases this is reflected by a roughly equal mix of the quarkonium component and the $B^{(*)} B^{(*)}$ component(s). $B_s^{(*)} B_s^{(*)}$ contributions, on the other hand, are almost negligible, because the corresponding threshold is more than $160 \, \text{MeV}$ above.

For higher resonances the meson-meson components start to dominate and the quarkonium component is somewhere between $8\% \ldots 35\%$. The corresponding widths tend to be sizable and the resonances are rather unstable. Thus, it is not surprising that these resonances are mostly meson-meson states.

For resonances above $\approx 11.025 \, \text{GeV}$, which is the threshold of one negative and another positive heavy-light meson, the quarkonium components start to increase again. We interpret this, however, rather as a consequence of our neglect of decay channels in that energy region than as a solid and meaningful physics result. We note again that also energy levels above $\approx 11.025 \, \text{GeV}$ should not be trusted or at least be taken with extreme caution, as e.g.\ discussed already in Section~\ref{SEC852}.

\begin{figure*}
	\includegraphics[width=0.48\textwidth]{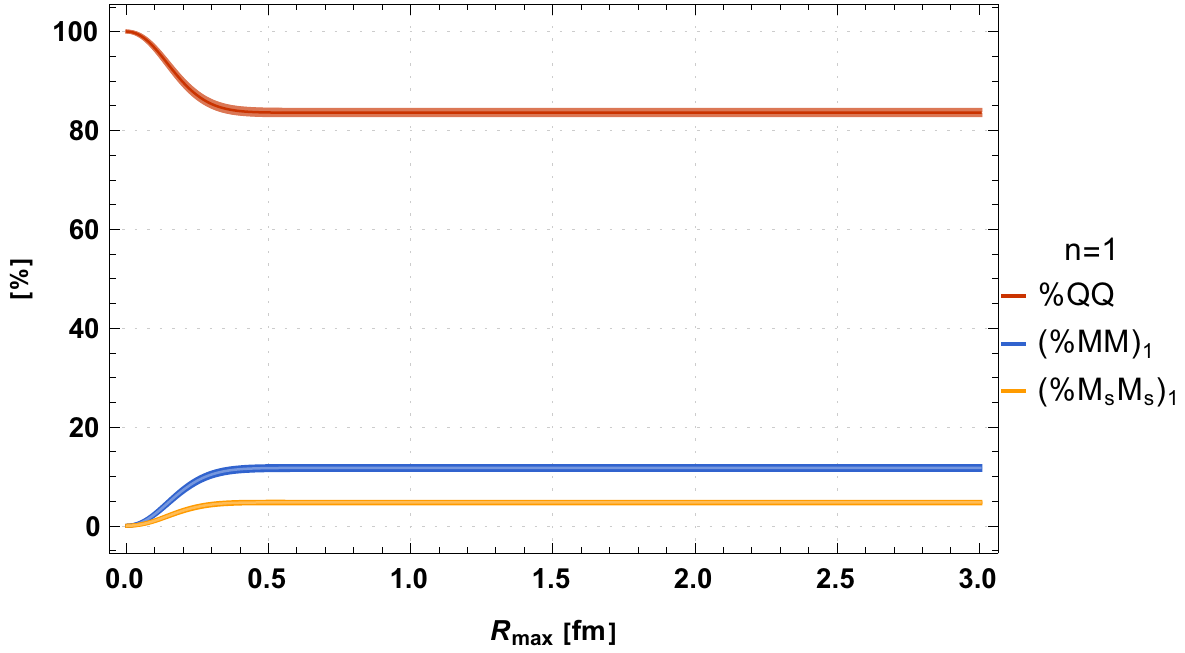}
	\includegraphics[width=0.48\textwidth]{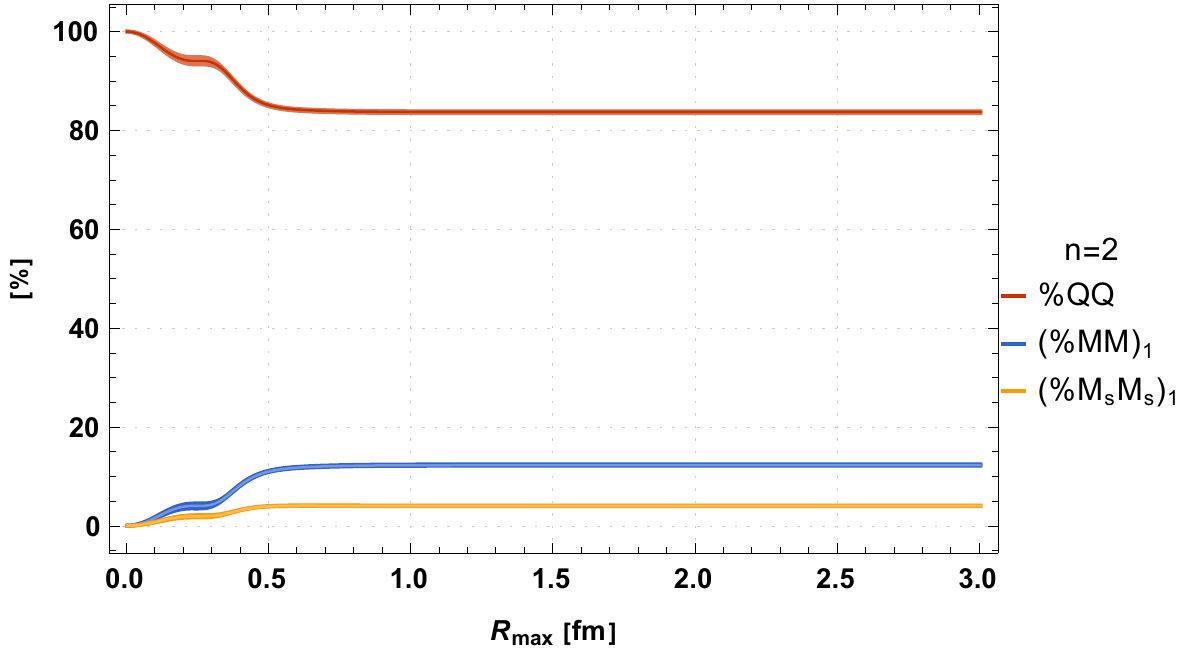}
	\includegraphics[width=0.48\textwidth]{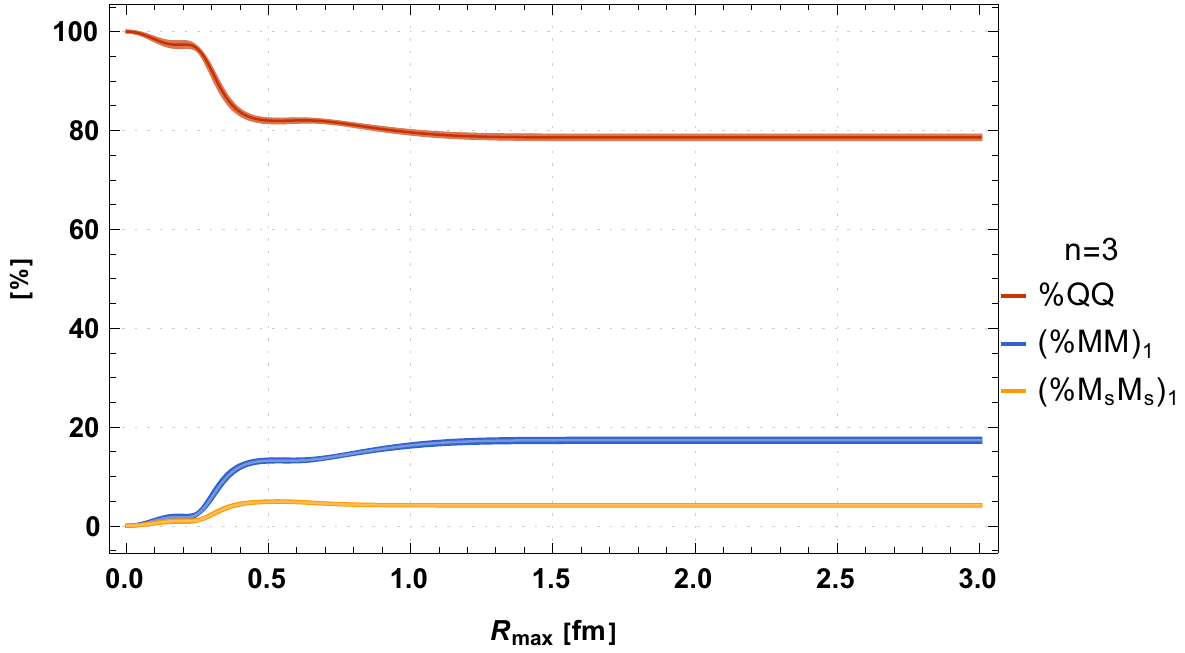}
	\includegraphics[width=0.48\textwidth]{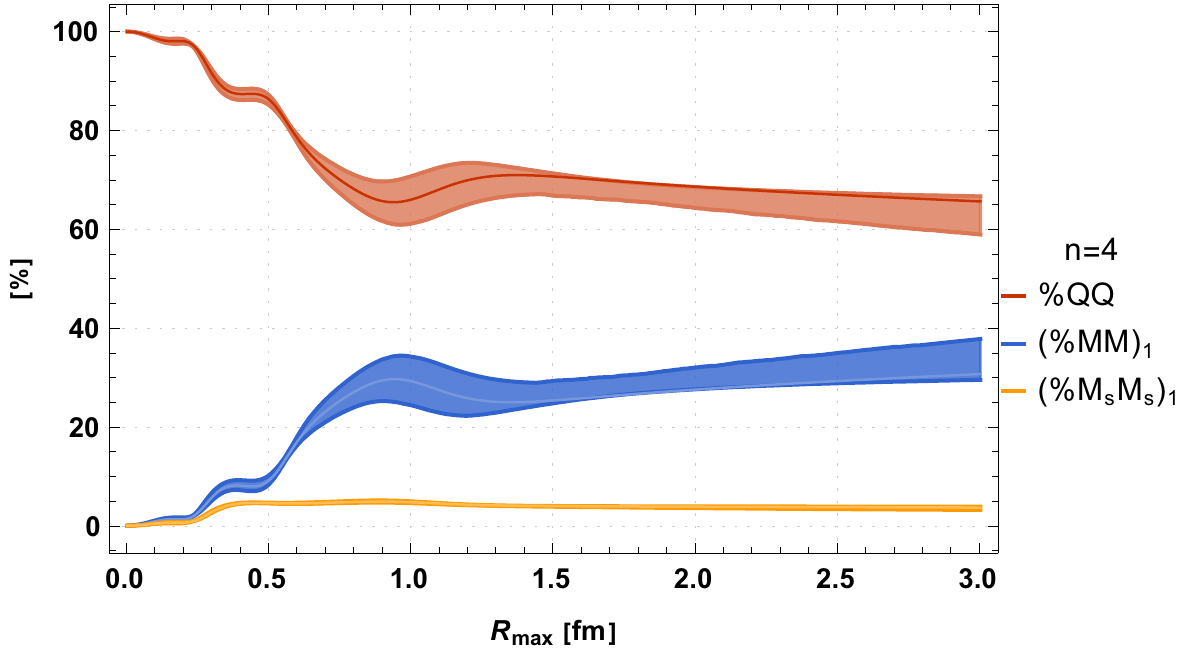}
	\includegraphics[width=0.48\textwidth]{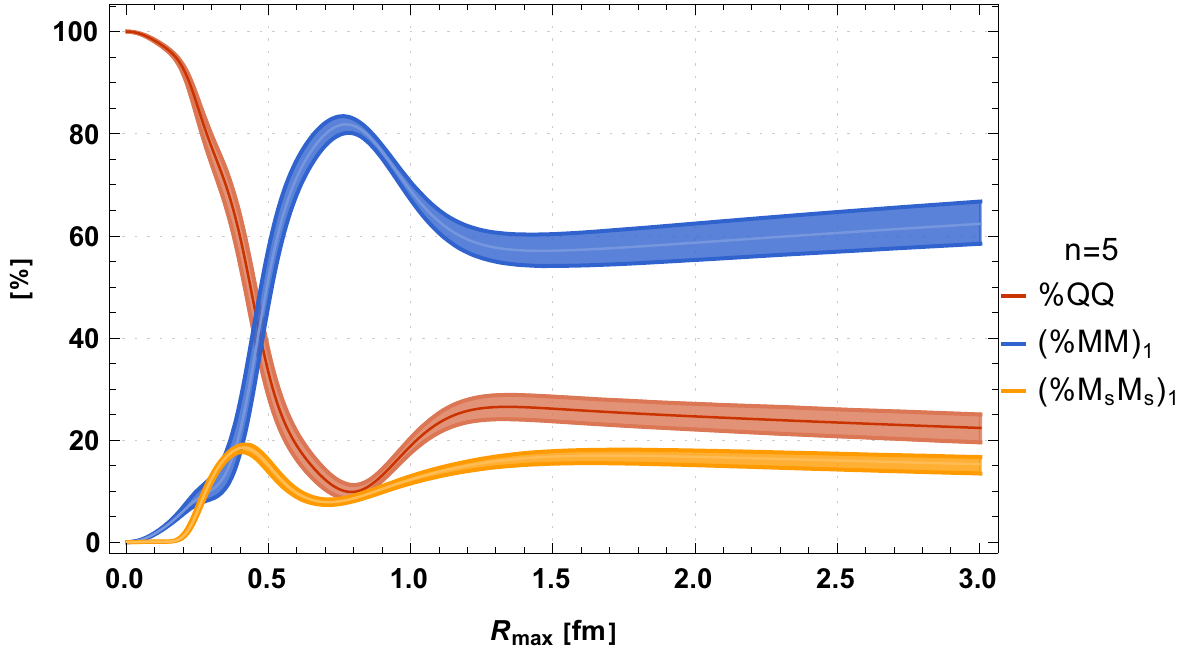}
	\includegraphics[width=0.48\textwidth]{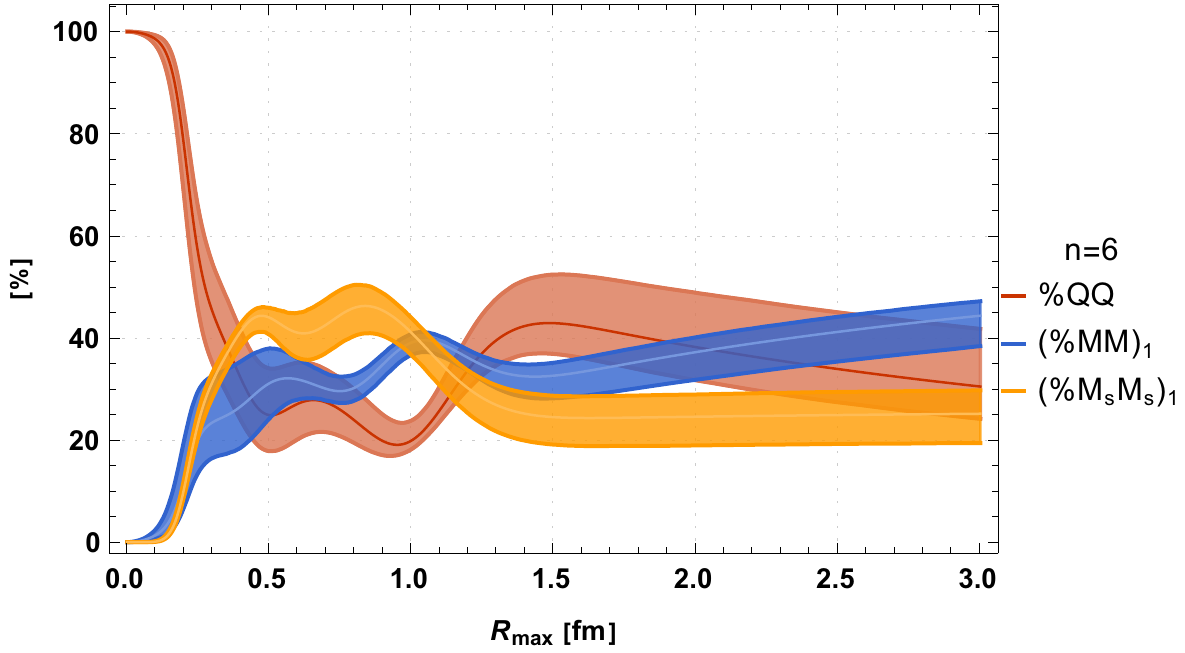}
	\includegraphics[width=0.48\textwidth]{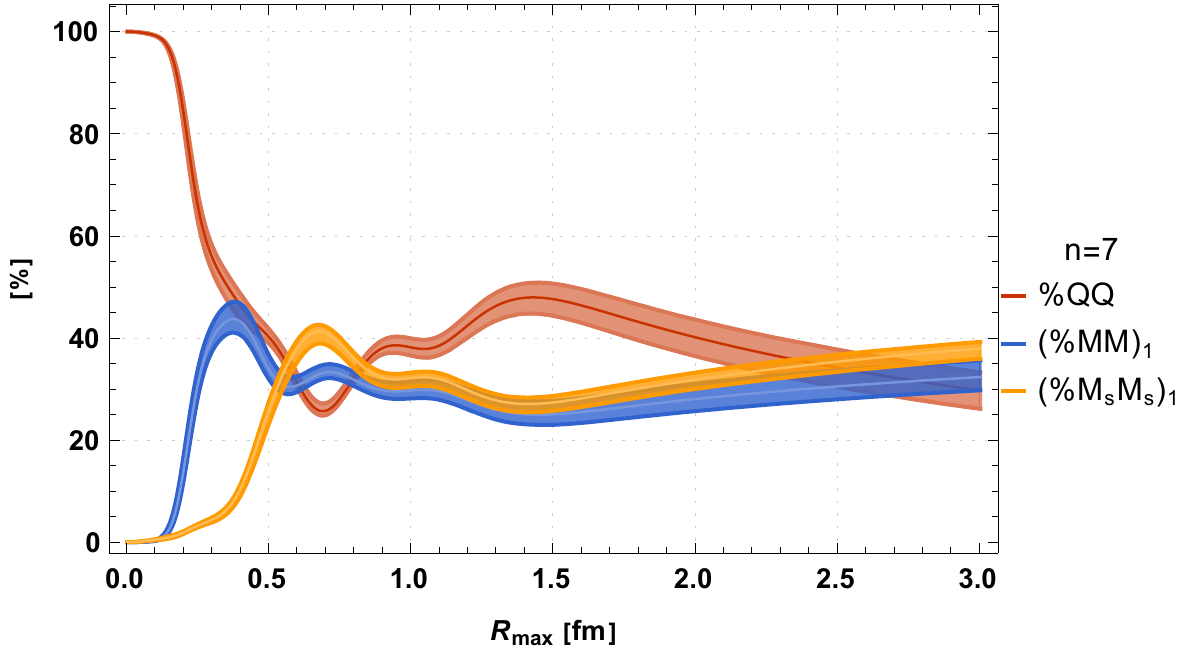}
	\caption{Percentages of quarkonium and of meson-meson pairs for $I = 0$ bottomonium with $\J^{PC}=0^{++}$ as functions of $R_{\textrm{max}}$.}
	\label{fig:percentages_J0}
\end{figure*}

\begin{figure*}
	\includegraphics[width=0.48\textwidth]{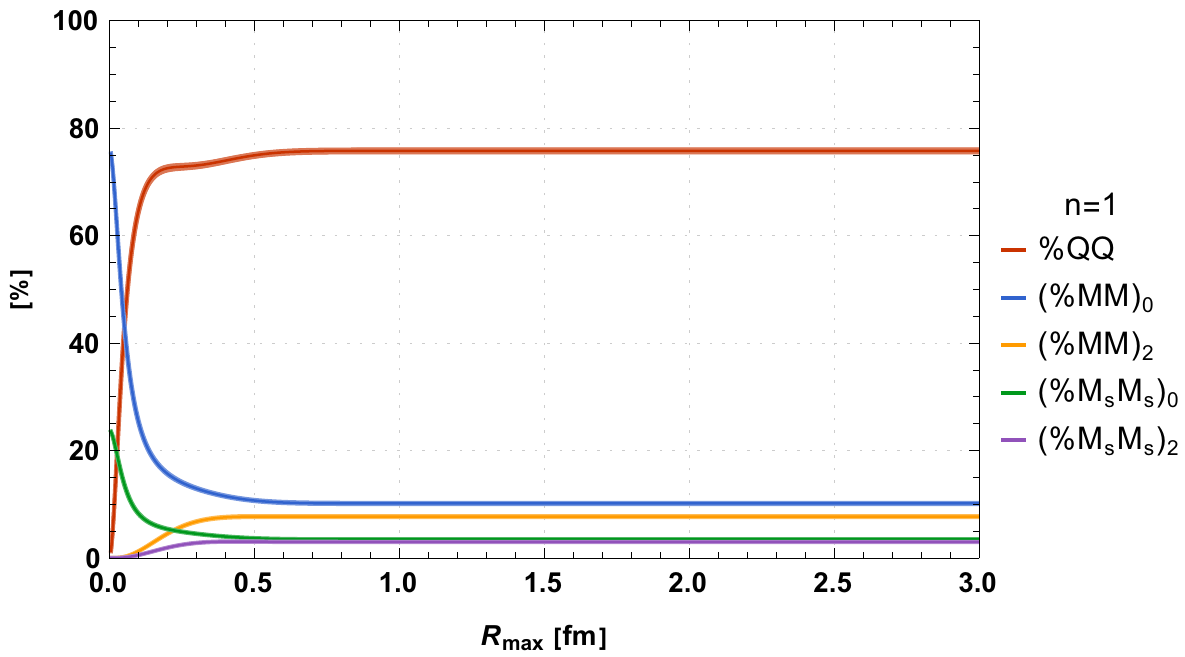}
	\includegraphics[width=0.48\textwidth]{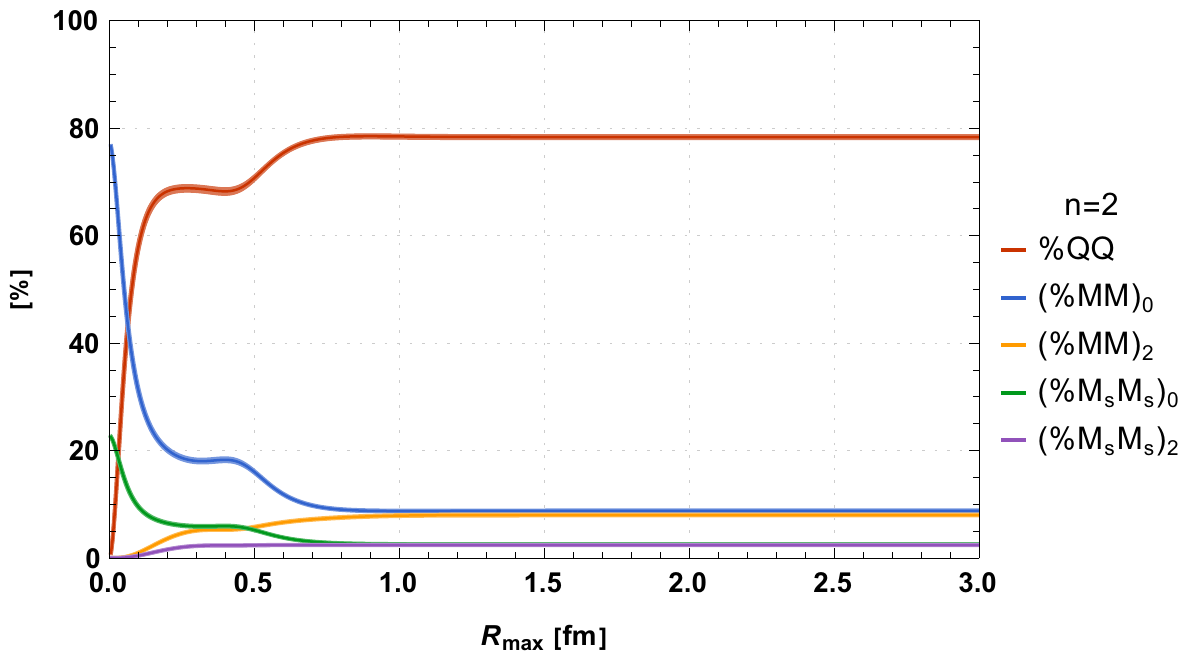}
	\includegraphics[width=0.48\textwidth]{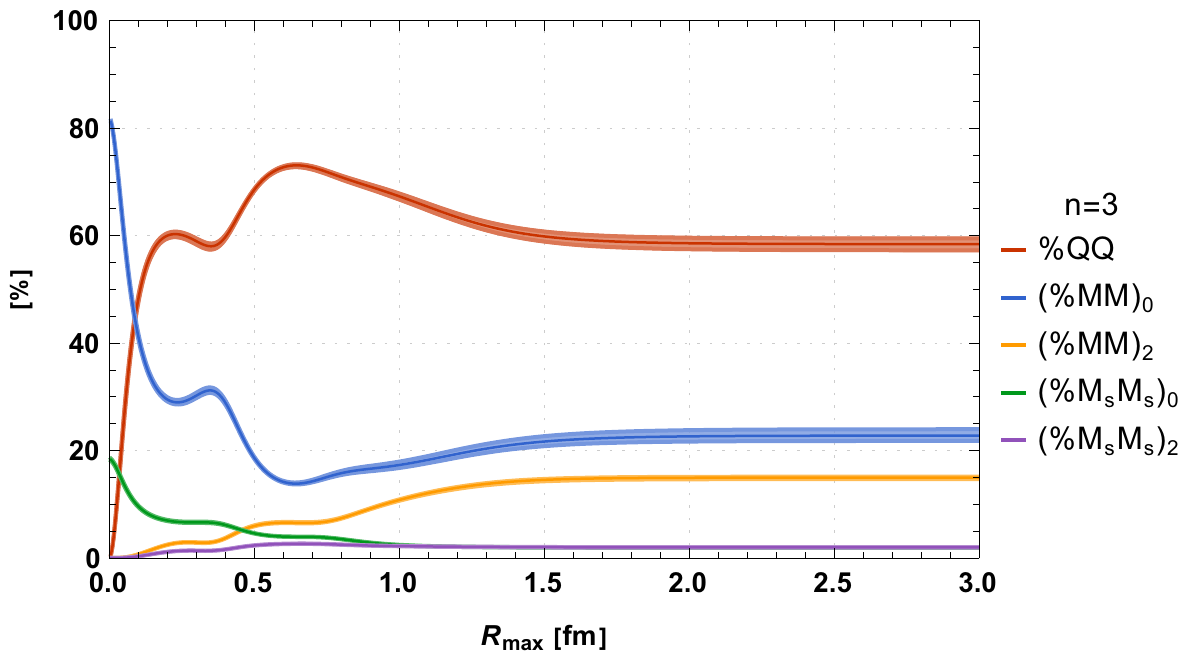}
	\includegraphics[width=0.48\textwidth]{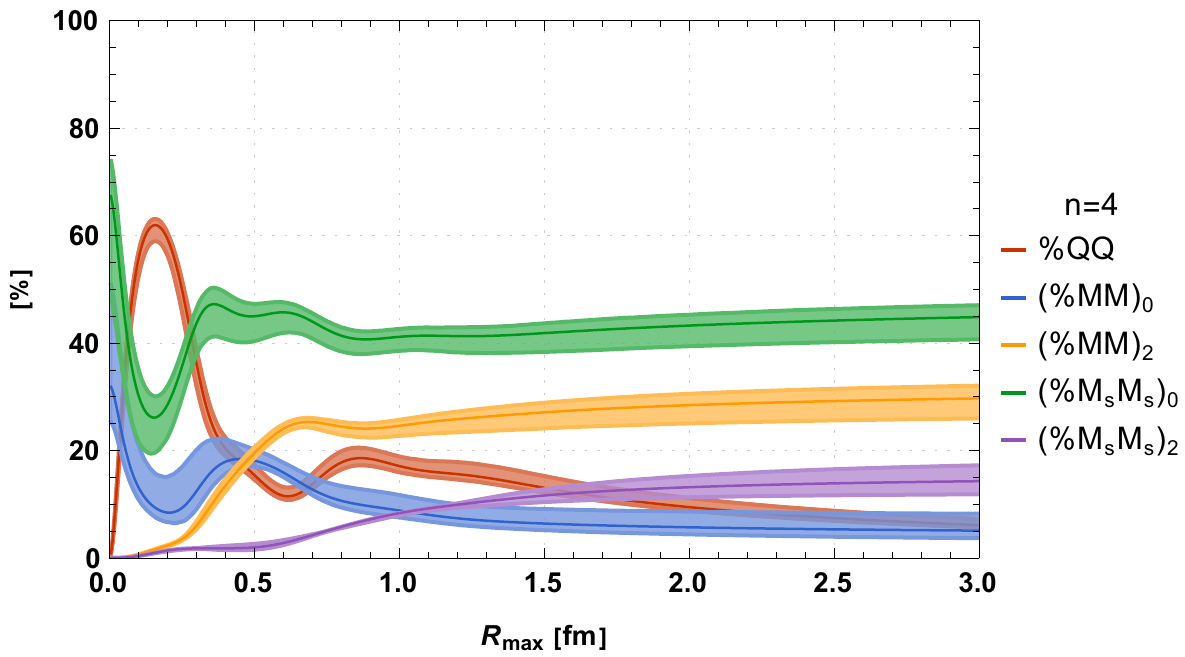}
	\includegraphics[width=0.48\textwidth]{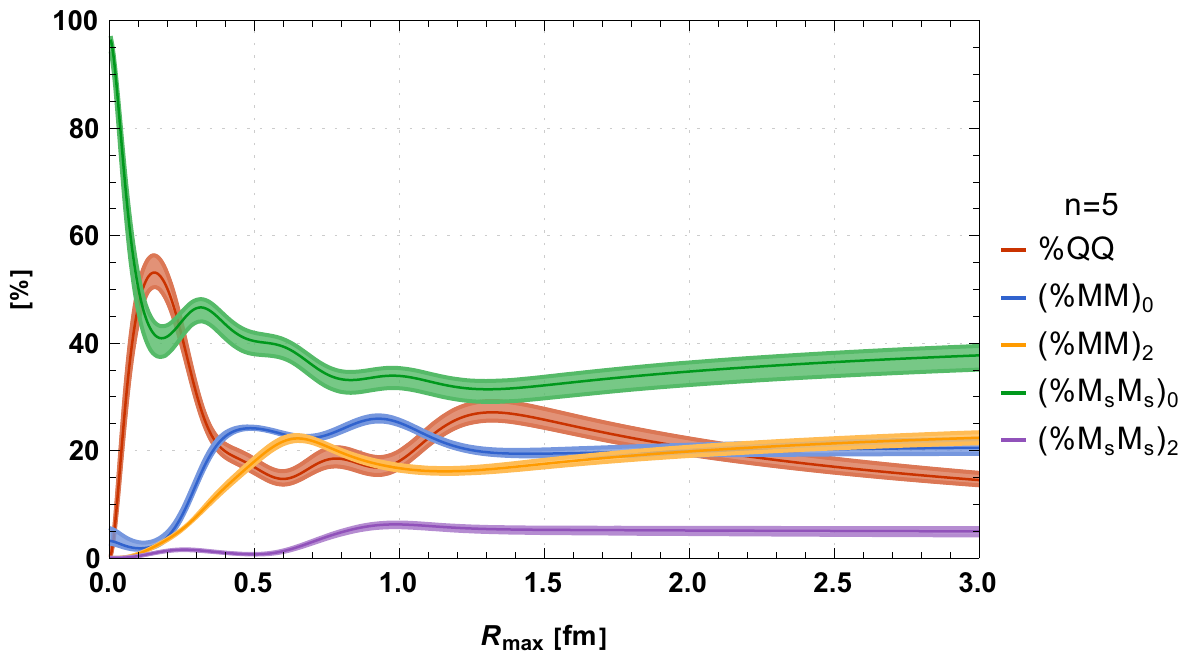}
	\includegraphics[width=0.48\textwidth]{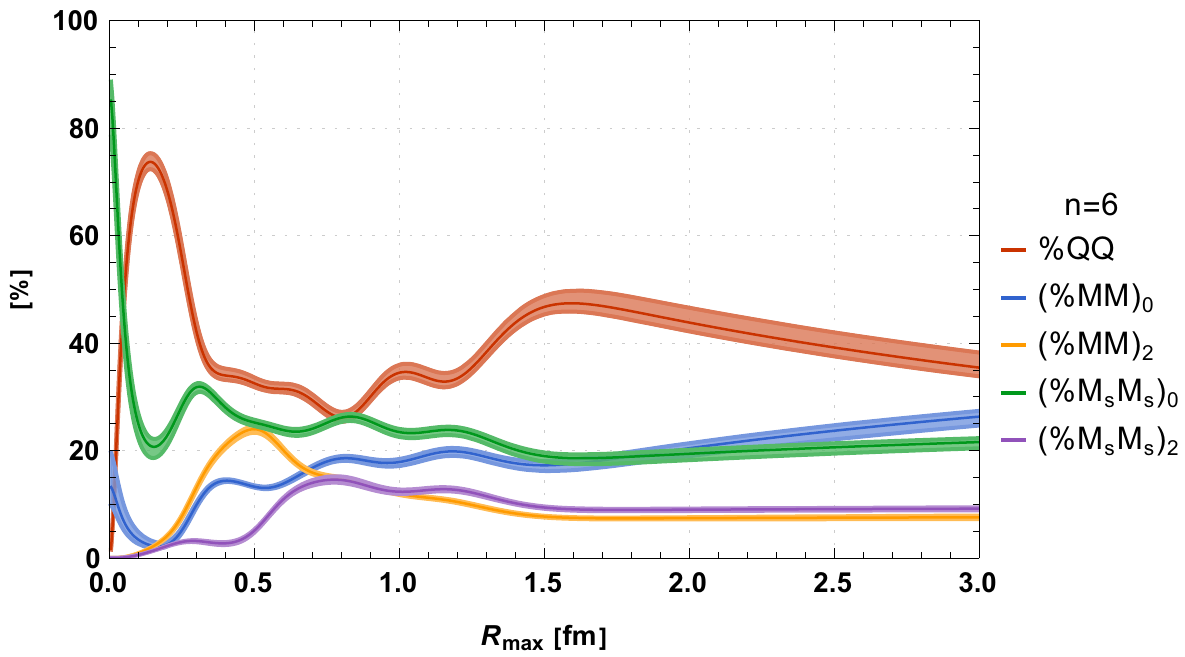}
	\caption{Percentages of quarkonium and of meson-meson pairs for $I = 0$ bottomonium with $\J^{PC}=1^{--}$ as functions of $R_{\textrm{max}}$.}
	\label{fig:percentages_J1}
\end{figure*}

\begin{figure*}
	\includegraphics[width=0.48\textwidth]{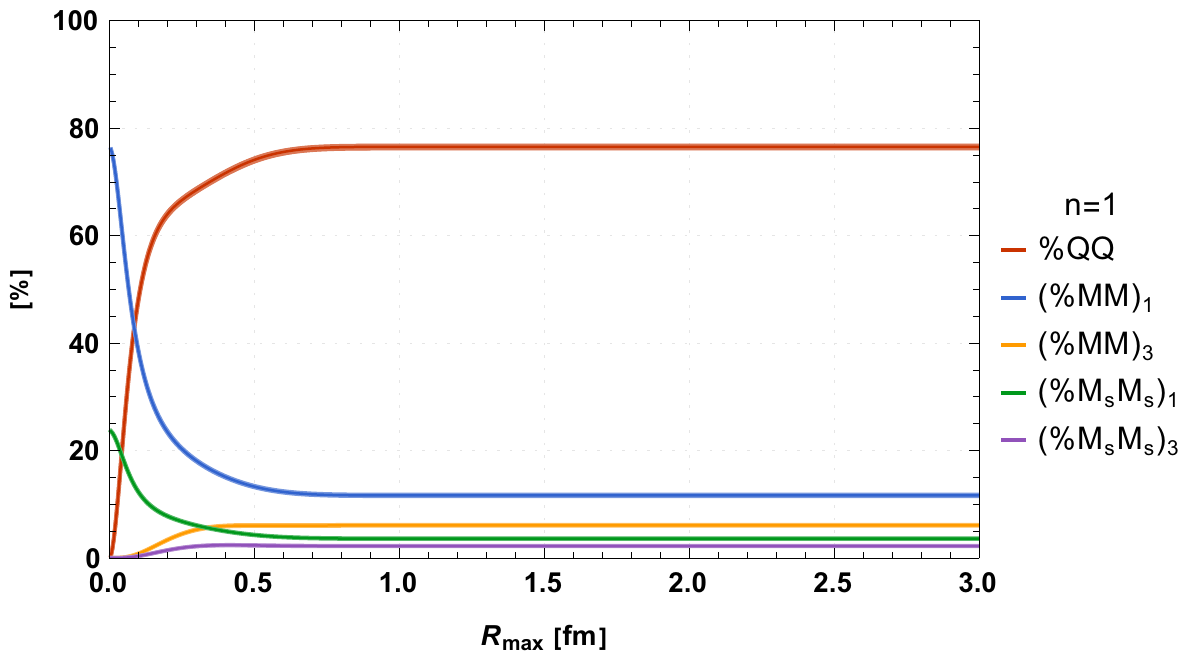}
	\includegraphics[width=0.48\textwidth]{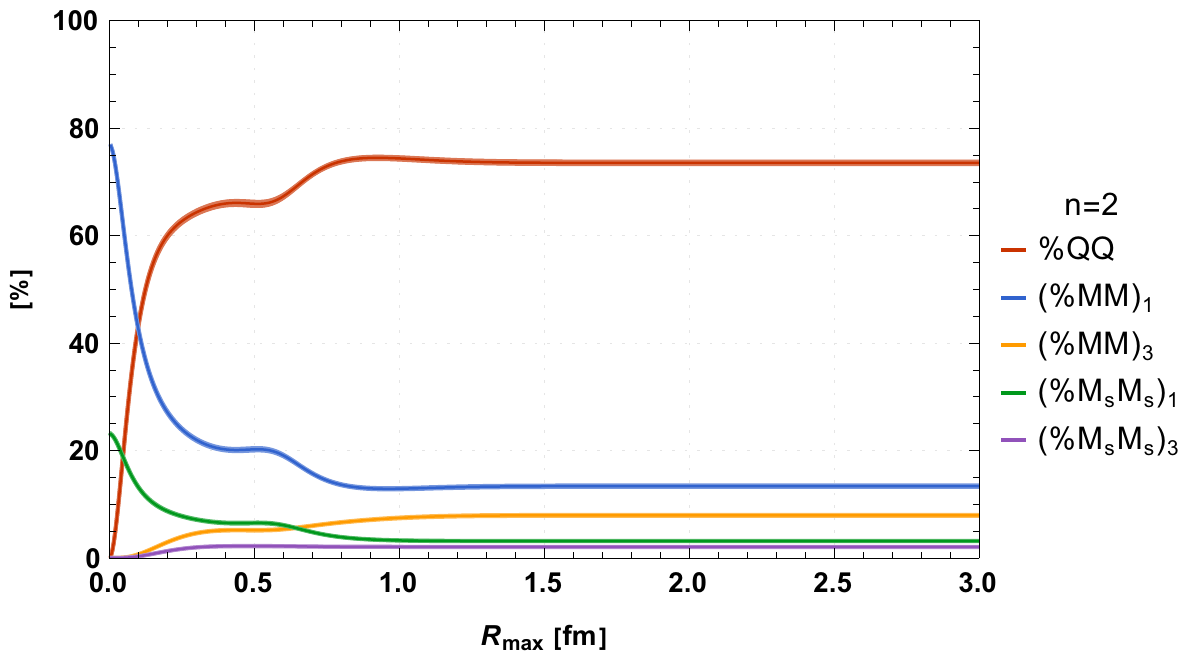}
	\includegraphics[width=0.48\textwidth]{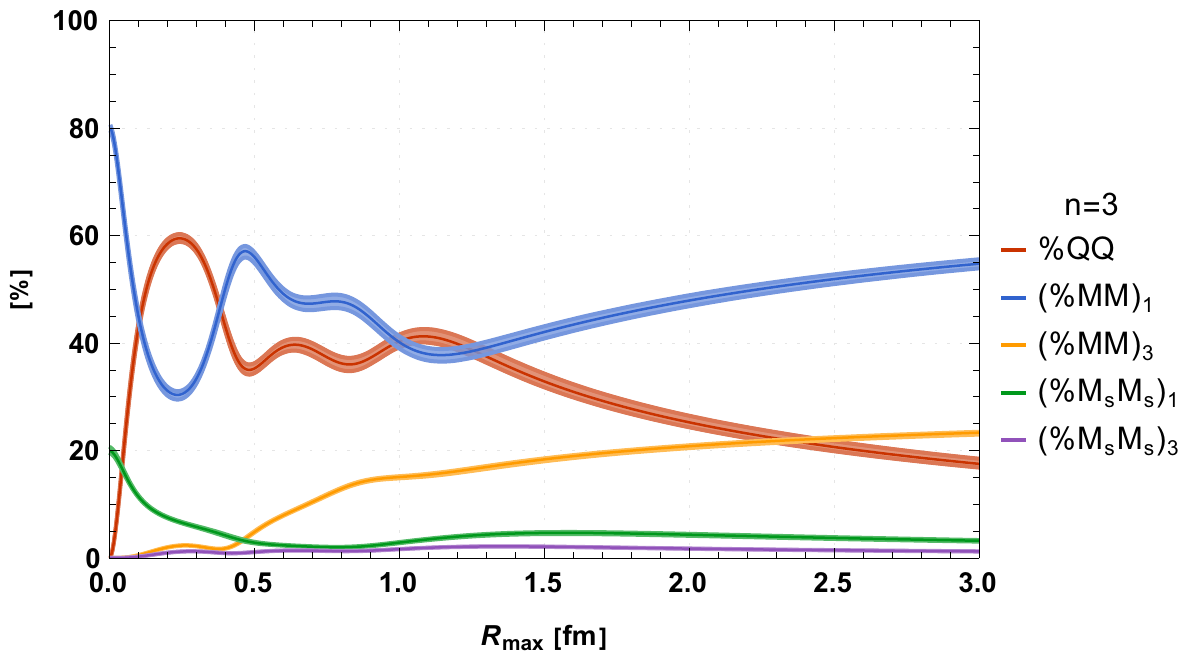}
	\includegraphics[width=0.48\textwidth]{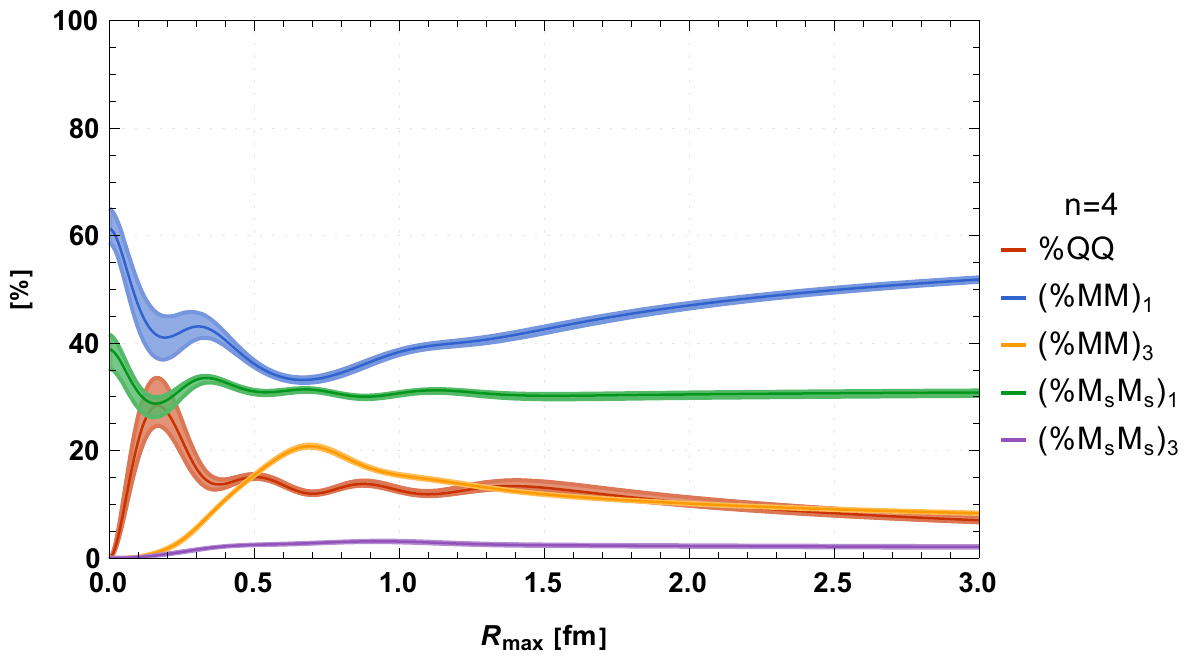}
	\caption{Percentages of quarkonium and of meson-meson pairs for $I = 0$ bottomonium with $\J^{PC}=2^{++}$ as functions of $R_{\textrm{max}}$.}
	\label{fig:percentages_J2}
\end{figure*}

\begin{figure*}
	\includegraphics[width=0.48\textwidth]{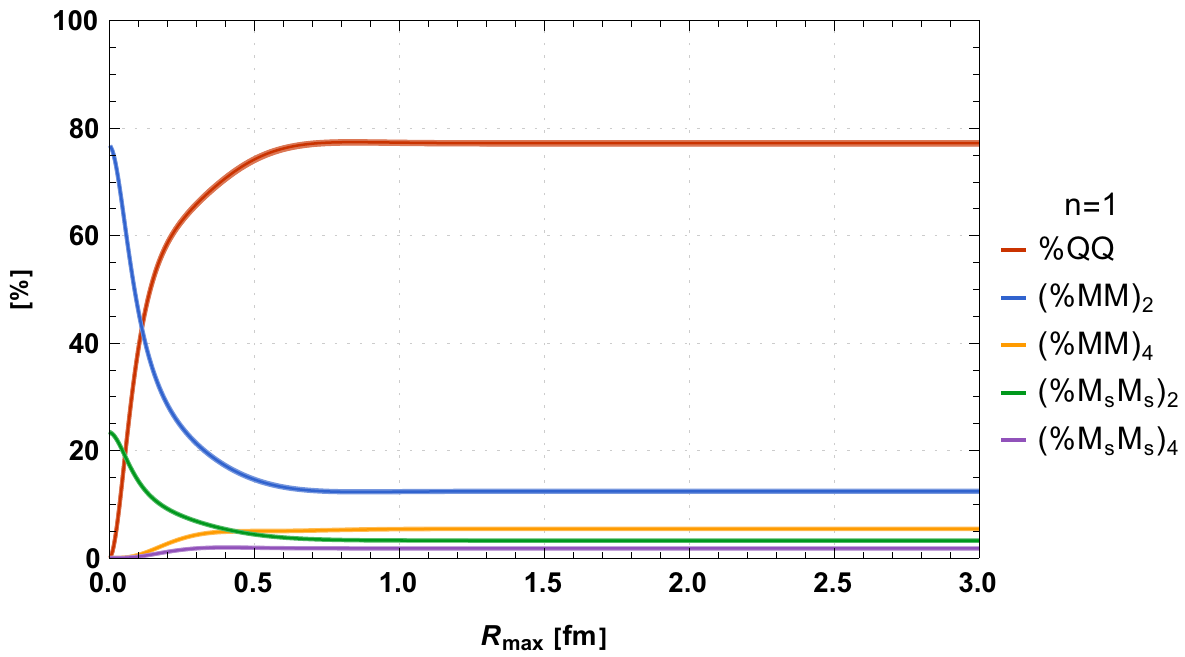}
	\includegraphics[width=0.48\textwidth]{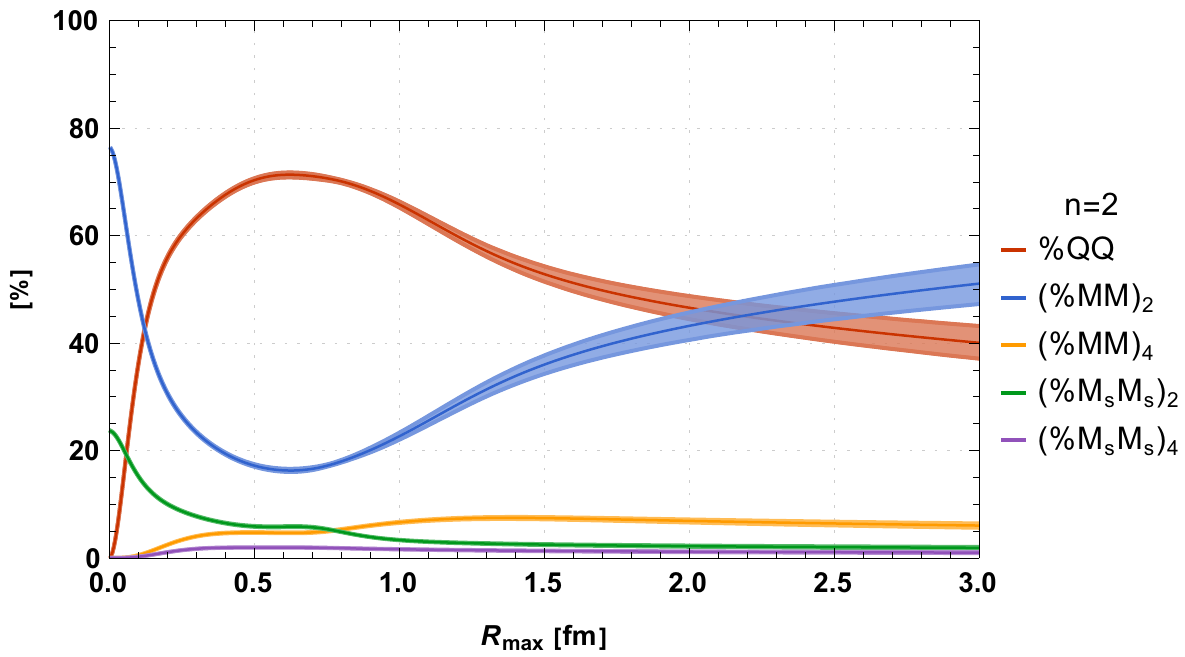}
	\includegraphics[width=0.48\textwidth]{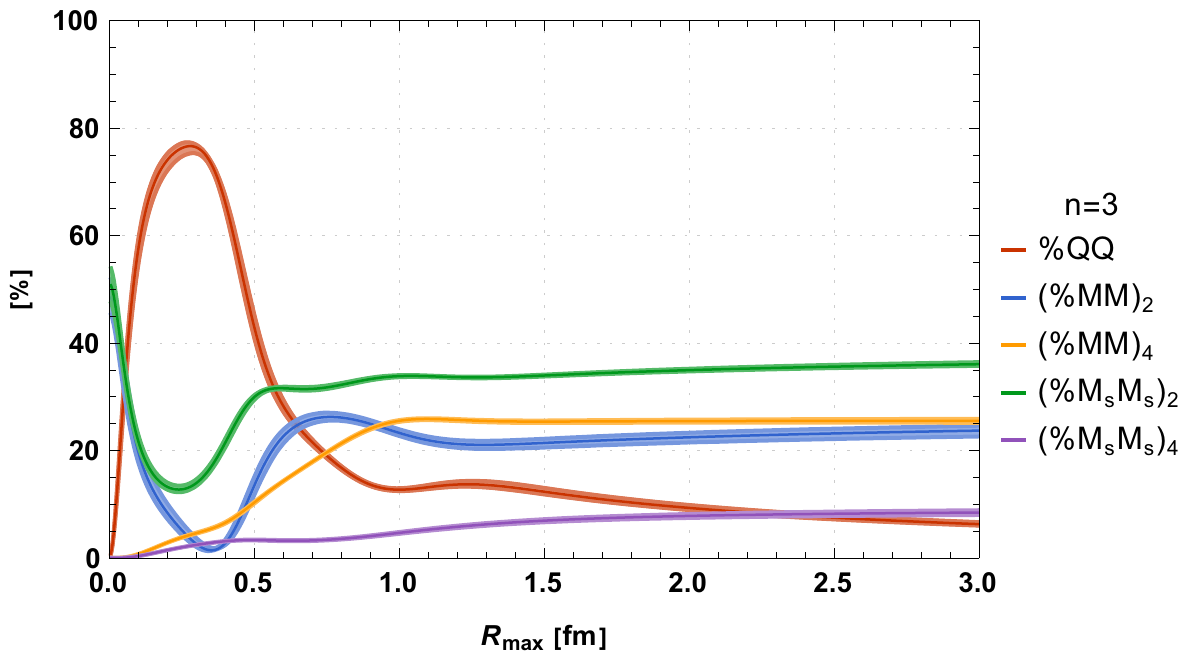}
	\includegraphics[width=0.48\textwidth]{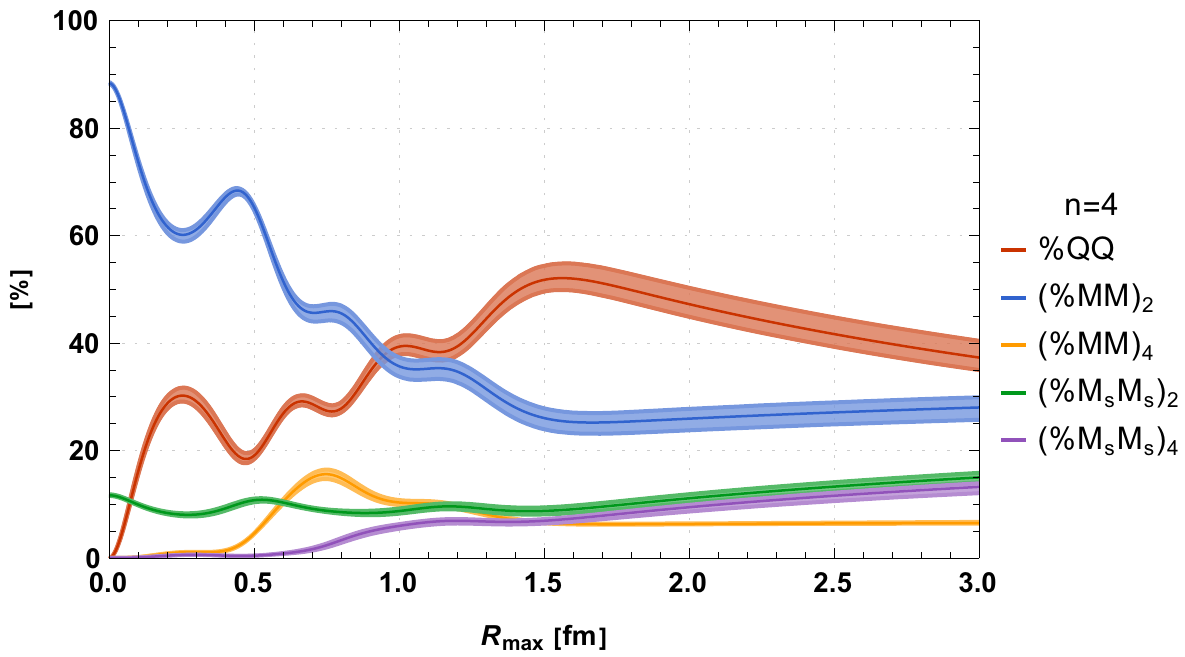}
	\caption{Percentages of quarkonium and of meson-meson pairs for $I = 0$ bottomonium with $\J^{PC}=3^{--}$ as functions of $R_{\textrm{max}}$.}
	\label{fig:percentages_J3}
\end{figure*}

\section{\label{SEC004}Conclusions and outlook}

We extended our previous work for $\J=0$ \cite{Bicudo:2019ymo,Bicudo:2020qhp} and derived a coupled channel Schrödinger equation for arbitrary $\J$ to investigate bottomonium bound states and resonances with $I=0$ using lattice QCD static potentials. For each $\J > 0$ we take into account five coupled channels and, thus, have developed an approach to study complicated resonances based on lattice QCD. We solved the coupled channel Schrödinger equation for $\J = 0,1,2,3$ and found multiple bound states and resonances for all values of $\J$. We also explored the structure of these states by computing their quarkonium and meson-meson percentages.

Our results for masses of bound states and resonances are consistent with experimentally observed states within expected errors, which arise from resorting to several approximations. 
We find several bound states in the sectors $\J = 0,1,2$, which all have a clear experimental counterpart. 

The resonance $\Upsilon(10753)$, which was recently found by Belle \cite{Abdesselam:2019gth} also appears in our $S$ wave spectrum as $\J = 0$, $n = 5$. See our last publication \cite{Bicudo:2020qhp}, where we discuss in detail that this is a meson-meson dominated state ,which can be classified as an $\Upsilon$ type crypto-exotic state. We now find another state with a comparable mass and decay width in our $D$ wave spectrum as $\J = 2$, $n = 3$, i.e.\ there are two distinct resonances very close in the spectrum. This is not surprising and can be observed already in constituent quark models, since $D$ wave quarkonium states have energies comparable to radially excited $S$ wave quarkonium states \cite{Godfrey:1985xj}.

We can confirm the interpretation of $\Upsilon(10860)$ as $\Upsilon(5S)$. Concerning $\Upsilon(11020)$ we find indications of corresponding resonances in both the $S$ wave sector ($\J = 0$, $n = 7$) as well as the $D$ wave sector ($\J = 2$, $n = 4$). We interpret these two resonances as two states as briefly discussed in the previous paragraph. We note that it would be very interesting, if experiments studying bottomonium could disentangle these two states.

In what concerns a possibly existing bottomonium state close to the $B^{(*)} B^{(*)}$ threshold, as counterpart to the $X(3872)$ charmonium state, we do not find one. However, we find a $\J=2$, $n=3$ state very close to the  $B_s^{(*)} B_s^{(*)}$ threshold, which could have similarities to $X(3872)$
\cite{Belle:2003nnu,CDF:2003cab} and its composite nature proposed by lattice QCD computations \cite{Padmanath:2015era} and by data analysis \cite{Kang:2016jxw}. This state has a meson-meson component significantly larger than its quarkonium component, $79 \%$ versus $21 \%$.

Our aim for the future is to reduce systematic errors as much as possible, to be able to reproduce the experimentally observed states not only on a qualitative level, but rather precisely with combined statistical and systematic errors of only a few MeV. To reduce the systematic errors, we plan to compute the necessary potentials $\VQQ$, $V_{\textrm{mix}}(r)$, $V_{\bar{M}M, \parallel}(r)$ and $V_{\bar{M}M, \perp}(r)$
 via lattice QCD explicitly, avoiding to perform algebraic operations on string breaking potentials from Ref.\ \cite{Bali:2005fu}. This should eliminate several sources of uncertainty and additionally create opportunities for new and interesting results:
\bie
\ie With up-to-date lattice results we might also obtain a more accurate scale setting, i.e.\ a more precise value for the lattice spacing $a$ or, equivalently, the Sommer parameter $r_0$. Moreover, we should be in a position to calibrate the bottom quark mass $m_b$ rather precisely and, thus, avoid using a value from quark models.

\ie There is a discrepancy between our lightest bound states and the corresponding experimental results. We expect that this can be resolved by a precise computation of the mixing angle $\theta(r)$, which has a strong impact on the potential matrix (see Section~II~A of Ref.\ \cite{Bicudo:2019ymo} for details).

\ie Related to the previous item is a precise determination of the mixing potentials between quarkonium and meson-meson pairs, providing quantitative first principles results for quantities estimated and used in a large number of quark models for many years \cite{Micu:1968mk,LeYaouanc:1972vsx,Kokoski:1985is,vanBeveren:1986ea,Bicudo:1989sj,Bruschini:2020voj}.

\ie The details of the short range and the long range parts of the entries of the potential matrix, if computed with sufficient precision, should be of interest for models based on hadron-hadron interactions.
\eie

Another significant source of systematic error is the neglect of heavy spin effects. To achieve the desired level of precision, we plan to include them using methods developed in a related project \cite{Bicudo:2016ooe} and possibly carry out a lattice QCD computation of $1 / m_b$ and $1 / m_b^2$ corrections at least for the confining potential $\VQQ$ (see e.g.\ Refs.\ \cite{Bali:1997am,Brambilla:2000gk,Pineda:2000sz,Brambilla:2004jw,Koma:2006si,Koma:2012bc}).

Finally, it might be interesting to include decay channels to a negative and a positive parity heavy-light meson pair, to make solid predictions up to the threshold of two positive parity mesons at around $11.525 \, \text{GeV}$. This would allow us to obtain information about several states, which are not yet measurable by experiments.

\begin{acknowledgements}

We acknowledge useful discussions with Gunnar Bali, Eric Braaten, Marco Cardoso, Francesco Knechtli, Vanessa Koch and Sasa Prelovsek.

P.B.\ and N.C.\ acknowledge the support of CeFEMA under the FCT contract for R\&D Units UIDB/04540/2020.
N.C.\ acknowledges the FCT contract SFRH/BPD/109443/2015.
L.M.\ acknowledges support by a Karin and Carlo Giersch Scholarship of the Giersch foundation.
M.W.\ acknowledges support by the Heisenberg Programme of the Deutsche Forschungsgemeinschaft (DFG, German Research Foundation) - project number 399217702.

Calculations on GPU servers of CeFEMA partly supported by NVIDIA were conducted for this research.
Calculations on the GOETHE-HLR and on the FUCHS-CSC high-performance computer of the Frankfurt University were conducted for this research. We would like to thank HPC-Hessen, funded by the State Ministry of Higher Education, Research and the Arts, for programming advice.
\end{acknowledgements}

\end{widetext}

\bibliographystyle{apsrev4-1}
\bibliography{literature}

\end{document}